%% file: sem.methoden.cm.tex
\documentclass[amssymb,twocolumn,showpacs,aps]{revtex4}
\usepackage{epsfig,amsmath,tabularx}
\begin{document}
\draft

\title{Soft Ellipsoid Model for Gaussian Polymer Chains}

\author{Frank Eurich and Philipp Maass}

\address{Fachbereich Physik, Universit\"at Konstanz, 78457
  Konstanz, Germany}

\date{\today}

\begin{abstract} 
  A soft ellipsoid model for Gaussian polymer chains is studied,
  following an idea proposed by Murat and Kremer [J.~Chem.~Phys.\ 
  \textbf{108}, 4340 (1998)]. In this model chain molecules are mapped
  onto ellipsoids with certain shapes, and to each shape a monomer
  density is assigned.  In the first part of the work, the
  probabilities for the shapes and the associated monomer densities
  are studied in detail for Gaussian chains. Both quantities are
  expressed in terms of simple approximate formulae. The free energy
  of a system composed of many ellipsoids is given by an
  intramolecular part accounting for the internal degrees of freedom
  and an intermolecular part following from pair interactions between
  the monomer densities. Structural and kinetic properties of both
  homogeneous systems and binary mixtures are subsequently studied by
  Monte-Carlo simulations. It is shown that the model provides a
  powerful phenomenological approach for investigating polymeric
  systems on semi-macroscopic time and length scales.
\end{abstract}

\maketitle

\input{introduction.cm}

\input{model.cm} 
\input{inputquantities.cm} 
\input{melt.cm}

\input{mixture.cm} 
\input{summary.cm} 
\input{appendix.cm}

\input{references.cm}

\end{document}

%% file: introduction.cm.tex
\section{Introduction}
\label{sec:introduction}

For many applications it is important to understand polymer dynamics on
semi-macroscopic time scales corresponding to configurational changes of
polymer chains on lengths scales comparable with or larger than the radius of
gyration $\bar R_G$. A problem of active current research in nano-technology,
for example, is the tailoring of thin polymer films on surfaces
\cite{Sung/etal:1996,Bruder+Brenn:1992}. Spontaneous phase separation
processes of incompatible polymer blends may be used to translate a chemical
pattern on the surface (as e.g.~produced by the micro-contact printing
technique) into a pattern of varying polymer compositions
\cite{Boltau/etal:1998}.

Numerical investigation of such problems by means of molecular
dynamics simulations of semi-microscopic bead-spring models is not
feasible. An overall configurational change of an entangled polymer
chain composed of typically $N\!=\!10^3$ monomers takes a time of
order $N^4\tau_v\!\approx\!1$s, where $\tau_v\!\approx\!1$ps is a
characteristic time for vibrational atomic motions. Changes of the
morphology (e.g.\ significant changes of phase domains during demixing
of incompatible polymer blends) then take several minutes, and this is
far beyond accessible time scales in computer simulations.

To overcome this problem, one needs to study coarse-grained models.
For describing spontaneous phase separation processes in polymer
blends the Cahn-Hilliard equation is often used. In more complex
situations involving hydrodynamic or elastic interactions, generalized
time-dependent Ginzburg-Landau equations may be employed. Exact
analytical treatments based on these equations are possible for the
linear short-time regime and can become rather rich for thin films
\cite{Fischer/etal:1998}. On the other hand, the exploration of the
large-scale dynamics by numerical means is hard even within
Ginzburg-Landau type treatments \cite{Kenzler/etal:2000}.
Nevertheless, numerous theoretical studies have been carried out in
the past, in particular for investigating the behavior in confined
geometries (for a recent review on this subject, see
\cite{Binder:1998}).

An alternative way to describe polymer dynamics on semi-macroscopic time
scales was suggested by Murat and Kremer \cite{Murat/Kremer:1998}. They
proposed a model, in which each polymer is described by an ellipsoid that can
change its shape, position and orientation. The probability for a particular
conformation follows from an intramolecular free energy functional. A
continuous monomer distribution is assigned to each ellipsoid for a given
conformation.  The ellipsoids can penetrate each other and their interaction
energy results from a pair potential. If the pair potential is given by a
delta function, the interaction energy between two ellipsoids is proportional
to the overlap integral of the respective monomer densities times an
interaction parameter $\hat\epsilon$ \cite{Murat/Kremer:1998}. A nice feature
of the ellipsoid model is that the input parameters
(cf.~Sec.~\ref{sec:inputquantities}) can be determined from more microscopic
models appropriate to describe the polymer system on short time scales. In
fact, in their original work Murat and Kremer considered a bead spring model
as a starting point and showed how the kinetics can be explored on long time
scales by employing the associated ellipsoid model.

In the present work we will present a simpler ansatz by considering
Gaussian chain molecules as underlying entities. We will show that the
associated ``Gaussian ellipsoid model'' (GEM) provides a powerful
phenomenological approach for describing the long-time kinetics of
interacting macromolecular systems.  Moreover, Gaussian chains serve
as a reference system for more realistic microscopic ``input models''.
Compared to these, the GEM has the following advantage: By including
self-interactions of the ellipsoids (see Sec.~\ref{sec:melt}), single
ellipsoids show Flory type behavior, $\bar R_G\!\sim\!N^{3/5}$, while
in dense systems with monomer concentration $c$ larger than the
overlap concentration $c_\star$ they exhibit a random walk type
behavior, $\bar R_G\!\sim\!N^{1/2}$. In contrast to the bead spring
model considered in \cite{Murat/Kremer:1998}, the $\bar
R_G\!\sim\!N^{1/2}$ scaling in dense systems is obtained without an
appropriate tuning of the interaction parameter $\hat\epsilon$ with the
chain length $N$. In fact, we will show that the GEM obeys the
well-known scaling relation $\bar R_G(c,N)\!\sim\!N^{3/5}f(cN^{4/5})$
describing the dependence of $\bar R_G$ on the polymerization degree
$N$ and monomer concentration $c$ \cite{deGennes:1979}. In this way
the GEM allows one to study diffusive dynamics on large length scales,
while accounting for the essential physical properties of polymeric
systems. 

From a technical point of view, we will show that the input quantities for the
GEM can be described by explicit approximate formulae (valid for
$N\!\gtrsim\!30$). This allows us to calculate the overlap integrals of
monomer densities analytically, thus speeding up the simulation considerably.
In addition, the explicit form of the input quantities is helpful for
employing the GEM in applications and may allow for analytical calculations in
the future.


%% file: model.cm.tex
\section{Soft Ellipsoid Model}
\label{sec:model}

The basic idea of the soft-ellipsoid-model is to map one polymer onto
one soft particle, that is to do an additional coarse--graining of a
microscopic model for polymers. \textit{Soft} means, that one
ellipsoid can overlap strongly with another. With respect to the setup
of the model outlined in this section, and the definitions and
notations involved, we follow closely the work of Murat and Kremer
\cite{Murat/Kremer:1998}.

Given a polymer with $(N\!+\!1)$ monomers at positions $\textbf{y}^{(k)}$
in a laboratory fixed coordinate system, $k\!=\!0,\ldots,N$, the
associated ellipsoid has the center of mass
$\textbf{r}=\sum_{k=0}^{N}\textbf{y}^{(k)}/(N\!+\!1)$ and its spatial
extension is determined by the eigenvalues $\textbf{S}\!=\!(S_1,S_2,S_3)$ 
of the radius of gyration tensor

\begin{equation} 
  \label{eq:radius.of.gyration.tensor}
  S_{\alpha\beta} = \frac{1}{N\!+\!1} \sum_{k=0}^{N}
  (y_{\alpha}^{(k)}-r_{\alpha}) (y_{\beta}^{(k)}-r_{\beta})\,,
\end{equation}

where $y_\alpha^{(k)}$ is the $\alpha$'th component of
$\textbf{y}^{(k)}$. Throughout this paper we denote spatial indices
by Greek letters and particle indices by Latin letters. The
orientation of the associated ellipsoid in space is given by the
orientation of the principal axis of $S_{\alpha\beta}$. Denoting the
corresponding rotation matrix by ${\cal R}$, the transformation from
coordinates $\textbf{y}$ of the laboratory fixed coordinate system to
coordinates $\textbf{x}$ of the ellipsoid's coordinate system is

\begin{equation}
\textbf{x}={\cal R}\,(\textbf{y}-\textbf{r})\,.
\end{equation}

Without loss of generality we choose
\begin{equation}
  \label{eq:Sa.condition}
  S_1 \ge S_2 \ge S_3\,,
\end{equation}
that means we order the eigenvalues so that the first coordinate in the
ellipsoid's coordinate system always refers to the largest axis, while the
third coordinate refers to the smallest axis.

Finally, a function $\varrho(\textbf{x};\textbf{S})$ is assigned to
each ellipsoid in its coordinate system, which specifies the mean
monomer density of the polymer for given eigenvalues $\textbf{S}$.
This means, that two ellipsoids $i,j$ with
$\textbf{S}_i\!=\!\textbf{S}_j$ have monomer densities
$\varrho(\textbf{x};\textbf{S}_{i}),\,
\varrho(\textbf{x};\textbf{S}_{j})$ that can be transformed into
each other by a spatial rotation. In the laboratory fixed coordinate
system the monomer density $\varrho_i'(\textbf{y})$ of polymer $i$ is
\begin{equation}
\varrho_i'(\textbf{y})=\varrho
({\cal R}_i(\textbf{y}-\textbf{r}_i);\textbf{S}_{i})\,.
\label{eq:trafo}
\end{equation}

The probability $P(\textbf{S})$ for a polymer to have eigenvalues
$\textbf{S}$ and the conditional monomer density
$\varrho(\textbf{x};\textbf{S})$ are the input quantities for the
soft-ellipsoid-model.  They have to be determined from a microscopic
model and their form for the GEM (Gaussian chains) is discussed in
Sec.~\ref{sec:inputquantities}.

The free energy functional of an ensemble of $M$ soft
particles in a volume $V$ is divided into an intramolecular part,
which accounts for the possible internal configurations, and an
intermolecular part, which describes the interaction between the
polymers,
\begin{equation}
F = F_{\text{intra}} + F_{\text{inter}}\,.
\label{eq:F}
\end{equation}
The intramolecular part is given by the probability $P(\textbf{S})$
\begin{equation}
  \label{eq:Fintra}
  F_{\text{intra}} = \sum_{i=1}^{M} F_{\text{intra}}^{(i)} =
  -k_{B}T \sum_{i=1}^{M} \ln P(\textbf{S}_{i})\, \text{,}
\end{equation}
where $k_{B}$ is the Boltzmann constant and $T$ is the
temperature. The intermolecular part is given by
\begin{eqnarray}
F_{\text{inter}} &=& \frac{1}{2} \sum_{i=1}^{M} \sum_{(j \neq i)}
  F_{\text{inter}}^{(ij)}+
\left(\frac{1}{2} \sum_{i=1}^{M}F_{\text{inter}}^{(ii)}\right)\,, 
\label{eq:Fintersum}\\
F_{\text{inter}}^{(ij)} &=& \int\hspace*{-0.1cm}
  d^3y\int\hspace*{-0.1cm}d^3z\,\,
  v(\textbf{y},\textbf{z})\, \varrho_{i}'(\textbf{y})\,
  \varrho_{j}'(\textbf{z})\label{eq:Finterv}\,,
\end{eqnarray}
We wrote ``$(j \neq i)$'' in eq.~(\ref{eq:Fintersum}), since one may or
may not include self-interaction terms. The effective pair potential
$v(\textbf{y},\textbf{z})$ between two monomers can include indirect
contributions mediated by solvent molecules. For the most simple
choice accounting for excluded volume effects \cite{Doi/Edwards},
$v(\textbf{y},\textbf{z})\!=\!\hat\epsilon\, b^3
\delta(\textbf{y}-\textbf{z})$ with $\hat\epsilon$ and $b^3$ being a
``contact energy'' and a ``contact volume'', respectively, the
interaction is given by the overlap integral of the monomer densities,
\begin{equation}
F_{\text{inter}}^{(ij)}
=\hat\epsilon\, b^3\int\hspace*{-0.1cm}d^3y\,\,
  \varrho_{i}'(\textbf{y})\, \varrho_{j}'(\textbf{y})\,.\label{eq:Finter}
\end{equation}
Equations (\ref{eq:F}--\ref{eq:Fintersum},\ref{eq:Finter}) fix the
thermodynamics of the model.

\begin{figure}[tbh]
  \begin{center}
    \epsfig{file=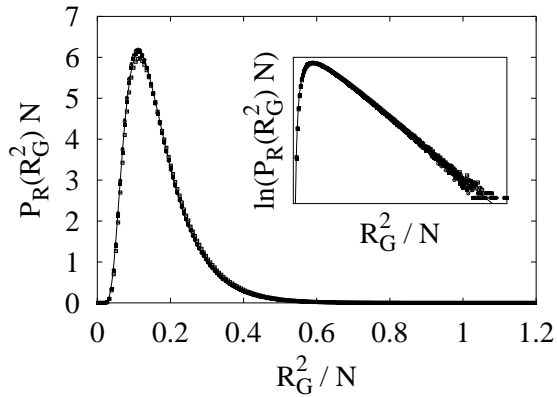,width=0.9\linewidth,angle=0}
  \end{center}
  \caption{Numerical results of the scaled probability function
    $P_{R}(R_{G}^{2})$ for $N = 30 (\square)$, $100 (\blacksquare)$,
    $300 (\circ)$ and $1000 (\bullet)$. The approximate function
    $P_{R}(R_{G}^{2})$ defined in the text (cf.  eq.~(\ref{fR(x)-eq})
    and Tab.~\ref{tab:probability.RG2.constants}) is drawn as a
    straight line. The inset shows the same data in semi-logarithmic
    form to prove the validity in the range of very small and very
    large $R_{G}^{2}$.}
  \label{fig:probability.RG2}
\end{figure}
Generalizations to polymer mixtures are straightforward
\cite{Murat/Kremer:1998}. In a binary polymer blend we have two types of
polymers $A$ and $B$. For simplicity we will assume that both types have the
same polymerization degree, $N_A\!=\!N_B\!=\!N$, and that a polymer of each
type interacts with a polymer of the same type with an interaction
$\hat\epsilon_{AA}\!=\!\hat\epsilon_{BB}\!=\!\hat\epsilon$, while polymers of
different types interact with
$\hat\epsilon_{AB}\!=\!\hat\epsilon\,(1+\delta)$ (see
Sec.~\ref{sec:mixture}).

To summarize, the external parameters defining the thermodynamic state
of a homogenous system are the polymerization degree $N$, the overall
monomer concentration $c\!=\!(N\!+\!1)M/V$, and the reduced
interaction strength $\epsilon\!\equiv\!\hat\epsilon/k_BT$. In a
binary blend we in addition have the mismatch interaction $\delta$ and
the fraction $f_A$ (or concentration $c_A$) of A polymers.

For describing kinetic properties we implement a discrete time
Monte-Carlo algorithm involving three different types of moves of a
randomly chosen ellipsoid \cite{Murat/Kremer:1998}: {\it (i)} A
translation, where the center of mass is displaced by a random vector
$\Delta\textbf{r}$, the components of which are drawn from a uniform
distribution in the interval $[-\Delta r_{\text{max}}/2, \Delta
r_{\text{max}}/2]$; {\it (ii)} A rotation of the ellipsoid with equal
probability in the steradian $4\pi$; {\it (iii)} A change of the
ellipsoid's size, where a triplet of random numbers $\Delta
S_{\alpha}$ uniformly distributed in the interval $[\max(-\Delta
S_{\text{max},\alpha}/2,-S_\alpha), S_{\text{max},\alpha}/2]$ is added
to the eigenvalues $\textbf{S}$. We restricted the changes $\Delta
S_\alpha$ to values that result only in positive $S_{\alpha}$.
Moreover, for a move according to rule {\it (iii)} it is possible that
the order of the principal axes changes because of
condition~(\ref{eq:Sa.condition}), which requires a simultaneous
update of the orientation of the ellipsoid. All possible moves are
attempted with the same probability and we define one Monte-Carlo step
(MCS) as $3\,M$ trials to change the state of an ellipsoid.

We note that the dynamics are purely diffusive and neglect advective
processes due to hydrodynamic flows. Moreover, the dynamics do not
capture entanglement effects, which are important on time scales
smaller than the disengagement time $\tau_D$ needed for a polymer to
diffuse over a distance of order $\bar R_G$ \cite{Doi/Edwards}.  As a
consequence, when adjusting the Monte-Carlo time step to real
experimental time scales, one should take care that the dependence of
$\tau_D$ on $N$ will generally be different in the simulations and the
real system.


%% file: inputquantities.cm.tex
\section{Gaussian Ellipsoid Model (GEM)} 
\label{sec:inputquantities}

The Gaussian chain is one of the best examined models for polymer
chains. In this model the distribution of bond vectors
$\Delta\textbf{y}$ between Kuhn segments is given by
\begin{equation}
  p(\Delta\textbf{y}) = \sqrt{\frac{3}{2\pi\, b^{2}}} 
  \exp\left(-\frac{3 \Delta\textbf{y}^2}{2 b^{2}}\right)\,\text{,}
\end{equation}
where for simplicity we have identified the root mean square bond length $b$
with the parameter specifying the contact volume in eq.~(\ref{eq:Finter}). We
define $b$ as our length unit, i.e.\ $b\!\equiv\!1$.

In the following we discuss the input quantities $P(\textbf{S},N)$ and
$\varrho(\textbf{x};\textbf{S},N)$. For clarity we have now explicitly
marked the dependence on $N$, but we take the freedom to suppress the
argument $N$, whereever it seems to be appropriate. Explicit
approximate formulae are determined for the input quantities based on
Monte Carlo simulations, as well as asymptotic expansions and
numerical results published in earlier work (see below). The
simulations are conducted following the procedure described in
\cite{Janszen/et.al.:1996}. Averages were performed over typically
$10^7$ different chain conformations.

\subsection{Probability distribution $P(\textbf{S})$}

The exact analytical form of the distribution function $P(\textbf{S},N)$ for
Gaussian chains is rather complicated. It can be presented in terms of
multiple integrals \cite{Wei/Eichinger:1990}, but an exact calculation from
the respective formulae is very cumbersome from the numerical point of view.
\begin{table}[!b]
  \caption{Parameter defining $P_{R}(R_{G}^{2})$ in
    eq.~(\ref{fR(x)-eq})} 
  \begin{center}
    \begin{tabular}{|c|l|}
      \hline 
      $n_{R}$          & 1.0      \\
      $a_{R}$          & 0.08020  \\
      $d_{R}$          & 1.842    \\
      $K_{0}(2 d_{R})$ & 0.015923 \\
      \hline
    \end{tabular}
  \end{center}
  \label{tab:probability.RG2.constants}
\end{table}
Useful explicit expressions are available only in the limit of large
eigenvalues $S_\alpha$ \cite{Wei/Eichinger:1990,Eichinger:1985}. To obtain a
convenient formula for practical computations, we first consider the simpler
distribution function $P_R(R_G^2,N)$ of the squared radius of gyration
\begin{equation}
\label{eq:sum.eigenvalues}
R_G^2 = {\rm tr}\, S_{\alpha\beta} = S_{1}+S_{2}+S_{3}\,,
\end{equation}
which can be expressed in terms of a single integral \cite{Yamakawa:1971}.
For large $N\to\infty$, $P_R(R_G^2,N)$ obeys the random walk scaling
\begin{equation}
P_R(R_G^2,N)\!\sim\!N^{-1}\tilde P_R(R_G^2/N), \hspace*{0.3cm}
\int_0^\infty \!du\, \tilde P_R(u)\!=\!1\,.
\label{pr-eq}
\end{equation}
The scaling function $\tilde P_R(u)$ behaves as $\tilde
P_R(u)\!\sim\!u^{\nu_0}\exp(-C_{0}/u)$ for $u\!\to\!0$ and $\tilde
P_R(u)\!\sim\!u^{\nu_\infty}\exp(-C_{\infty}\,u)$ for $u\!\to\!\infty$
\cite{Fixman:1962,Forsman/Hughes:1963} with $C_{0}$, $C_{\infty}$,
$\nu_0$, and $\nu_\infty$ being constants. Note that the first moment
of $\tilde P_R(u)$ must equal 1/6 to obtain the well known result
$\bar R_G\!\equiv\!\langle R_G^2\rangle^{1/2}\!\sim\!(N/6)^{1/2}$ in
the limit of large $N$.

In this section, $\langle\ldots\rangle$ denotes an average over an
ensemble of Gaussian chains, while in
Secs.~\ref{sec:melt},\ref{sec:mixture} it will also be used with
respect to an ensemble of interacting ellipsoids.  From the context it
will be clear, what kind of average is meant.

Since the exponential terms dominate the behavior of $\tilde P_R(u)$
for small and large $u$ we make the ansatz
\begin{equation}
\tilde P_R(u)\!\cong\! 
\frac{u^{-n_{R}}\,(a_{R}d_{R})^{n_{R}\!-\!1}}{2\,K_{n_R\!-\!1}(2\,d_{R})} 
\exp\left(-\frac{u}{a_{R}}-d_{R}^{2}\frac{a_R}{u}\right)\,,
\label{fR(x)-eq}
\end{equation}
where $K_{n_R\!-\!1}(.)$ denotes the modified Bessel function of
$(n_R\!-\!1)$th order accounting for proper normalization. For
simplicity we restrict $n_R$ to half--integers, but it is not supposed
to equal $-\nu_0$ or $-\nu_\infty$. Since the form of the maximum
turns out to depend sensitively on $n_R$, we determined $n_R$ by a
least square fit of (\ref{fR(x)-eq}) to the Monte Carlo data in the
region given by the full width at half maximum. After determining
$n_R\!=\!1$ by this procedure, we calculated the constants $a_R$ and
$d_R$ from the first two moments of $\tilde P_R(u)$ (see Appendix A
and Tab.~\ref{tab:probability.RG2.constants}).

Fig.~\ref{fig:probability.RG2} shows that the scaling (\ref{pr-eq}) is
well obeyed for $N\!\gtrsim\!30$ \cite{small.N} and that
eq.~(\ref{fR(x)-eq}) provides an excellent approximation to the
simulated data.  Moreover, higher moments $\langle R_G^{2m}\rangle$,
$m\!=\!2,3,\ldots$ calculated by using eq.~(\ref{fR(x)-eq}) deviate by
less than one per cent from the exact result (see Appendix A).

\begin{table}[!b]
  \caption{Parameter defining $P_{\alpha}(S_{\alpha})$ in
    eq.~(\ref{eq:prod.PS})}
  \begin{center}
     \begin{tabular}{|c|c|c|c|}
      \hline
      $S_{\alpha}$ & $S_{1}$ & $S_{2}$ & $S_{3}$ \\ 
      \hline
      $n_{\alpha}$ & 1/2 & 5/2 & 4 \\ 
      $a_{\alpha}$ & 0.08065 & 0.01813 & 0.006031 \\
      $d_{\alpha}$ & 1.096 & 1.998 & 2.684 \\
      $K_{n_{\alpha}-1}(2 d_{\alpha})$ & 0.094551 & 0.0144146 &
      0.0052767 \\
      \hline
    \end{tabular}
  \end{center}
  \label{tab:probability.Sa.constants}
\end{table}
\begin{figure}[tbh]
  \begin{center}
    \epsfig{file=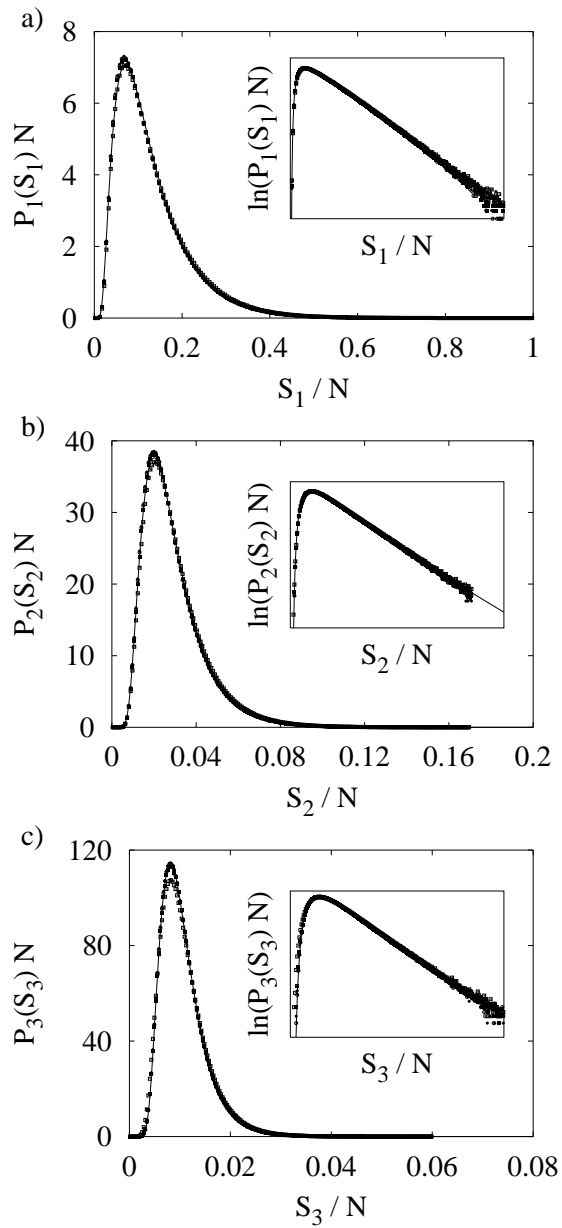,width=0.9\linewidth,angle=0}
  \end{center}
  \caption{Comparison of Monte-Carlo results for the scaled
    probability functions $P_{\alpha}(S_{\alpha})$ for $N = 30
    (\square)$, $100 (\blacksquare)$, $300 (\circ)$ and
    $1000(\bullet)$ with the approximate formulae defined in
    eq.~(\ref{eq:prod.PS}) and
    Tab.~\ref{tab:probability.Sa.constants}. The insets show the
    respective data in semi-logarithmic form again.}
  \label{fig:probability.Sa}
\end{figure}
To find a useful approximate formula for $P(\textbf{S},N)$, we first consider
the distributions
\begin{equation}
  \label{eq:Sa.heuristic}
  P_{\alpha}(S_\alpha,N) = \int d^{2}S_{\beta}\,
  P(\textbf{S},N)\; , \quad \beta \neq \alpha \,,
\end{equation}
of the single eigenvalues $S_\alpha$, $\alpha\!=\!1,2,3$. In view
of eq.~(\ref{eq:sum.eigenvalues}) these can be expected to behave
similar to $P_R(R_G^2,N)$.  Using the random walk scaling for large
$N$, we write $P_\alpha(S_\alpha,N)\!\sim\!N^{-1}\tilde
P_\alpha(S_\alpha/N)$, $\int_0^\infty\! du\,\tilde P_\alpha(u)\!=\!1$,
with $\tilde P_\alpha(.)$ as in eq.~(\ref{fR(x)-eq}), except for the
fact that the parameters $n_R$, $a_R$, and $d_R$ are replaced by
$n_\alpha$, $a_\alpha$, and $d_\alpha$. These parameters have been
\begin{figure}[!t]
  \begin{center}
    \epsfig{file=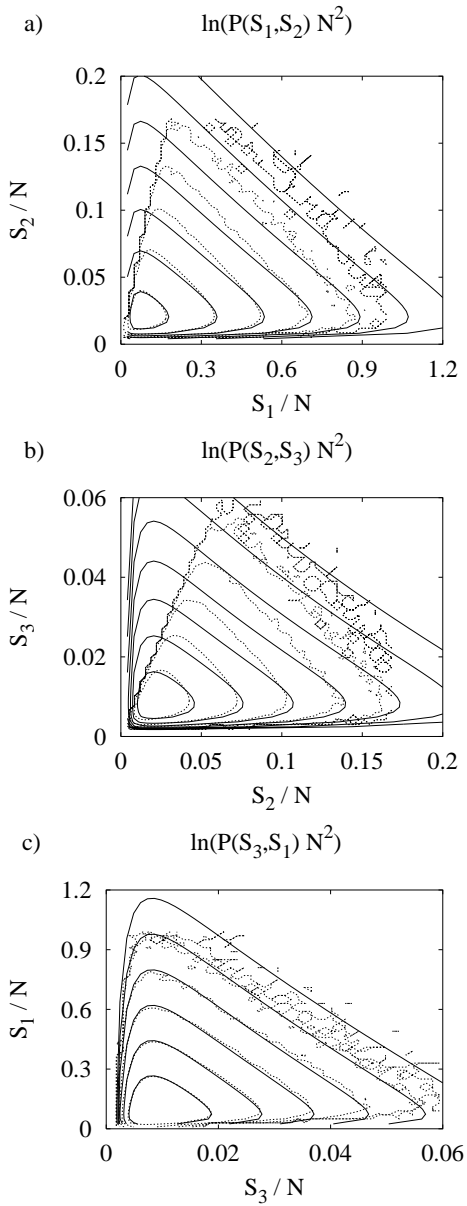,width=0.75\linewidth,angle=0}
  \end{center}
  \caption{Scaled contour plots showing the comparison of $\ln
    (P_{\alpha}(S_{\alpha}) P_{\beta}(S_{\beta}) N^{2})$ (straight
    lines) with numerical data of $\ln P(S_{\alpha},S_{\beta})$
    (dashed lines). The isolines are drawn at function values of
    $10^{n}$, where n is an arbitrary integer number. The agreement of
    the separation ansatz eq.~(\ref{eq:prod.PS}) with the numerical
    data is quite good especially in the region of large values, which
    have the highest statistical weight. However the separation ansatz
    neglects condition eq.(\ref{eq:Sa.condition}). For large values of
    $S_{\alpha}$ the Monte-Carlo data become noisy due to insufficient
    statistics.}
  \label{fig:PS.comparison}
\end{figure}
determined analogous to the procedure described above (for details see
Appendix A) and they are listed in
Tab.~\ref{tab:probability.Sa.constants}. Again, there is an excellent
agreement between these functions and the simulated data, see
Fig.\ref{fig:probability.Sa}.

\begin{table}[!b]
  \vspace*{0.5cm}
  \caption{Results for the correlation coefficients to test ansatz
      eq.(\ref{eq:prod.PS}).}
  \begin{center}
    \begin{tabular}{|c|l|}
      \hline
      $(\langle S_{1}S_{2} \rangle - \langle S_{1} \rangle \langle
      S_{2} \rangle) / \langle S_{1}S_{2} \rangle$ & 0.054 \\ 
      $(\langle S_{2}S_{3} \rangle - \langle S_{2} \rangle \langle
      S_{3} \rangle) / \langle S_{2}S_{3} \rangle$ & 0.059 \\
      $(\langle S_{3}S_{1} \rangle - \langle S_{3} \rangle \langle
      S_{1} \rangle) / \langle S_{3}S_{1} \rangle$ & 0.017 \\
      \hline
    \end{tabular}
  \end{center}
  \label{tab:correlation.coefficients}  
  \vspace*{0.5cm}
\end{table}
Next, in order to obtain the combined probability density $P(\textbf{S},N)$,
we neglect correlations among the $S_\alpha$, and write
\begin{eqnarray}
  \label{eq:prod.PS}
&& P(\textbf{S},N)=\prod_\alpha P_\alpha(S_\alpha,N)
=\prod_\alpha \frac{1}{N}\,\tilde P_\alpha\left(\frac{S_\alpha}{N}\right)\\
&&=\prod_\alpha
\frac{S_{\alpha}^{-n_{\alpha}}\,(a_{\alpha}d_{\alpha}N)^{n_{\alpha}\!-\!1}} 
     {2\, K_{n_{\alpha}\!-\!1}(2\,d_{\alpha})} 
\exp\left(-\frac{S_{\alpha}}{a_{\alpha}N} -
  d_{\alpha}^{2}\frac{a_{\alpha} N}{S_{\alpha}}\right)\,.\nonumber
\end{eqnarray}
A useful test of the accuracy of eq.~(\ref{eq:prod.PS}) is
to check whether the correlation coefficients
\begin{equation}
\gamma_{\alpha\beta}^{(S)}\equiv
\frac{\langle S_\alpha S_\beta \rangle - \langle S_\alpha
  \rangle \langle S_\beta \rangle}{\langle S_\alpha S_\beta
  \rangle}\,,\hspace*{0.3cm} \alpha \neq \beta \;.
\end{equation}
are much smaller than one. The results summarized in
Tab.~\ref{tab:correlation.coefficients} show that all
$\gamma_{\alpha\beta}^{(S)}$ are less than 6\%. We also checked
whether the asphericity \cite{Rudnick/Gaspari:1986}
\begin{equation}
\langle A \rangle= \frac{\langle R_{G}^{2}\rangle - 
3\langle S_1S_2 +
  S_1S_3 + S_2S_3\rangle}{\langle R_{G}^{2} \rangle} 
\end{equation}
of the ellipsoids is significantly changed compared to the true value. From
the simulations we find $\langle A \rangle\!\cong\!0.527$, while from
eq.~(\ref{eq:prod.PS}) it follows $\langle A \rangle\!=\!0.542$, which
amounts to a relative error of about 3\%.

A more sensitive test is the direct comparison of the combined probability
density
\begin{equation}
  P(S_{\alpha},S_{\beta}) = \int_0^\infty d S_{\gamma}\, P(\textbf{S}) \,,
\hspace*{0.3cm} \gamma \neq \alpha, \beta
\end{equation}
with $P_\alpha(S_{\alpha})P_\beta(S_{\beta})$. As shown in
Fig.~\ref{fig:PS.comparison}, the agreement between the ansatz
(\ref{eq:prod.PS}) and the true behavior obtained from the simulations
remains good, except for the regions where $P(S_{\alpha},S_{\beta})$
becomes extremely small.  We plotted the logarithm of
$P(S_{\alpha},S_{\beta})$ in Fig.~\ref{fig:PS.comparison}, since we
are interested in determining $F_{\text{intra}}\!\propto\!\ln
P(\textbf{S})$. The main restriction of (\ref{eq:prod.PS}) is that it
does not obey the condition (\ref{eq:Sa.condition}), i.e.\ 
$P(\textbf{S})\!=\!0$ for $S_2,S_3\!<\!S_1$ and $S_3\!<\!S_2$.
However, this poses no severe problem, because the difference in the
three scales $\langle S_1\rangle : \langle S_2 \rangle : \langle S_3
\rangle = 12.07 : 2.717 : 1$, cf. eg. Tab.~\ref{tab:Sa.moments},
ensures that the location of the maximum of $P(\textbf{S})$ is well
separated from the region, where eq.~(\ref{eq:prod.PS}) breaks down.
Due to the exponential decrease $P_\alpha\!\propto\!\exp(-C_\alpha
/S_\alpha)$ for small $S_{\alpha}$, $P$ is very small in the region,
where it should be exactly zero.

\subsection{Monomer density $\varrho(\textbf{x};\textbf{S})$}

\begin{table}[!b]
  \caption{Moments of the eigenvalues $\textbf{S}$ as determined by
      Monte-Carlo simulation. The table displays results for chains of
    length $N=1000$ averaged over $10^{7}$ different realizations. The
    data for the radius of gyration $R_{G}$ are given for the sake of
    comparison. The deviations from the exact values for $N
    \rightarrow \infty$ are due to statistical errors and finite chain
    lengths.}
  \begin{center}
    \begin{tabular}{|cl|cl|}
      \hline
      $\langle R_{G}^{2} \rangle/N$ & 0.16685 & $\langle R_{G}^{4}
      \rangle/N^{2}$ & 0.035302 \\
      \hline
      $\langle S_{1} \rangle/N$ & 0.12758   & $\langle S_{1}^{2}
      \rangle/N^{2}$ & 0.023009 \\
      $\langle S_{2} \rangle/N$ & 0.028708  & $\langle S_{2}^{2}
      \rangle/N^{2}$ & 0.0010300 \\
      $\langle S_{3} \rangle/N$ & 0.010568  & $\langle S_{3}^{2}
      \rangle/N^{2}$ & 0.00013143 \\
      \hline
      $\langle S_{1}S_{2} \rangle/N^{2}$ & 0.0038714  & $\langle S_{1}^{3}
      \rangle/N^{3}$ & 0.0055805  \\
      $\langle S_{2}S_{3} \rangle/N^{2}$ & 0.00032239 & $\langle S_{2}^{3}
      \rangle/N^{3}$ & $4.6206\,10^{-5}$ \\
      $\langle S_{3}S_{1} \rangle/N^{2}$ & 0.0013717  & $\langle S_{3}^{3}
      \rangle/N^{3}$ & $1.9338\, 10^{-6}$ \\
      \hline
    \end{tabular}
  \end{center}
  \label{tab:Sa.moments}
  \vspace{0.2cm}
\end{table}
\begin{figure}[!t]
  \begin{center}
    \epsfig{file=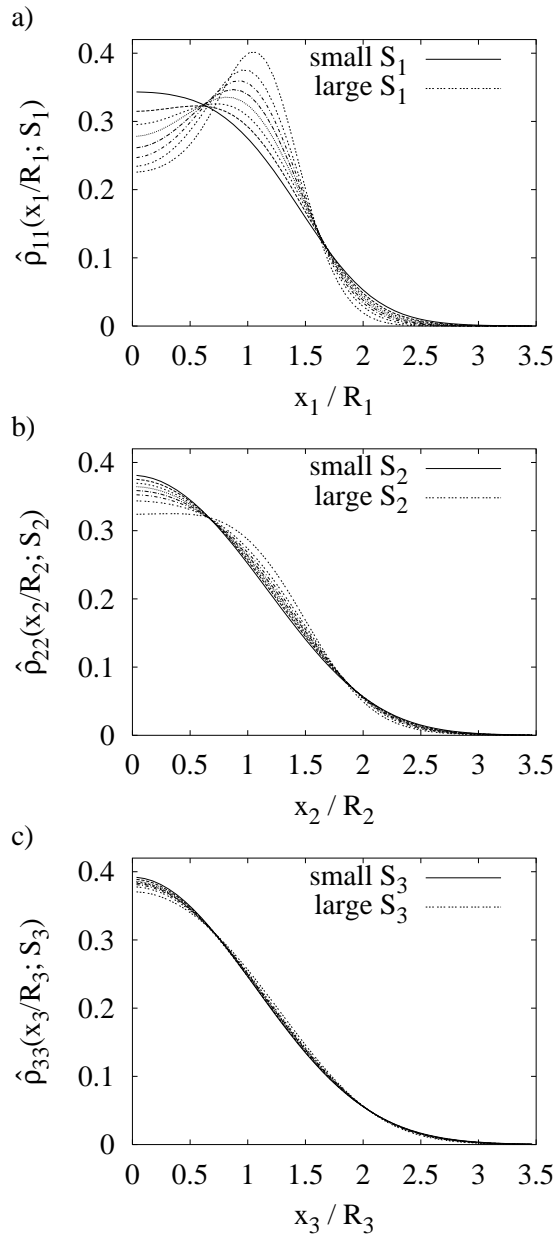,width=0.9\linewidth,angle=0}
  \end{center}
  \caption{One-dimensional conditional monomer density
    $\hat\varrho_{\alpha\beta}(x_{\alpha}/R_{\alpha};S_{\beta})$,
    scaled with $R_{\alpha}=\sqrt{S_{\alpha}}$, to demonstrate the
    breakdown of condition (\ref{eq:selfsimilar.density}) in a strict
    sense. The curves were sampled for chain-lengths $N=1000$.}
  \label{fig:conditional.density}
\end{figure}
For $\varrho(\textbf{x};\textbf{S},N)$ no exact analytical expression is
available.  Since $\varrho(\textbf{x};\textbf{S},N)$ is a function of many
variables, various simplifications are necessary to allow for an efficient
computation of the interaction terms in eq.~(\ref{eq:Finter}). Murat and
Kremer \cite{Murat/Kremer:1998} proposed a self-similar monomer distribution
($R_\alpha\!=\!S_\alpha^{1/2}$),
\begin{equation} 
  \label{eq:selfsimilar.density}
  \varrho(\textbf{x};\textbf{S},N) = \frac{N\!+\!1}{\prod_\alpha R_\alpha}\,
\tilde\varrho(\textbf{x}/\textbf{R})\,,
\end{equation}
where we introduced the short-hand notation
$\tilde\varrho(\textbf{x}/\textbf{R})\!\equiv
\!\tilde\varrho(x_{1}/R_{1},x_{2}/R_{2},x_{3}/R_{3})$.  In contrast to
\cite{Murat/Kremer:1998}, we included the factor
$[(N\!+\!1)/\prod_\alpha R_\alpha]$ in
eq.~(\ref{eq:selfsimilar.density}) due to the normalization
\begin{equation}
  \int d^3x\,\varrho(\textbf{x};\textbf{S},N)=N\!+\!1\,.
\end{equation}

While the self-similarity assumption seems to be plausible, we will next show
that it is not strictly valid. To this end we consider the function
$[\prod_\alpha R_\alpha]\,
\varrho(\textbf{x};\textbf{S},N)/(N\!+\!1)$ in terms of
the scaled variable $\textbf{u}\!\equiv\!\textbf{x}/\textbf{R}$, i.e.\ we
consider
\begin{equation}
\hat\varrho(\textbf{u};\textbf{S},N)\!\equiv\!
\frac{\prod R_\alpha}{N\!+\!1}\,\varrho(R_1 u_1,R_2 u_2, R_3 u_3;
\textbf{S},N)\,.  
\end{equation}
If the scaling
(\ref{eq:selfsimilar.density}) would hold true, this function would equal
$\tilde\rho(\textbf{u})$, and hence any averaging over regions in
$\textbf{S}$--space would leave it invariant. In particular we define
\begin{eqnarray}
\hat\varrho_{\alpha\beta}(u_{\alpha};S_\beta,N)&\equiv&\nonumber\\[0.1cm]
&&\hspace*{-2cm}\int_{-\infty}^\infty
\hspace*{-0.05cm}\prod_{\mu\neq\alpha}du_\mu 
\int_0^\infty \hspace*{-0.15cm}\prod_{\nu\neq\beta}d S_\nu\, P(\textbf{S},N)\,
\hat\varrho(\textbf{u};\textbf{S},N)\,,
\end{eqnarray}
that means the scaled monomer density with respect to one principal axis
$\alpha$ under the condition that one arbitrary eigenvalue
corresponding to the axis $\beta$ is given. These are nine different
functions, and the Monte-Carlo results for some of them are shown in
Fig.~\ref{fig:conditional.density} for $N\!=\!1000$. For other
$N\!\gtrsim\!30$ we find almost exactly the same functions, that means
the scaling with $N$ is well obeyed.  In order to obtain the functions
displayed in the figure, we divided the $S_{\alpha}$--axis in eight
different intervals, delimited at positions $S_\beta^{(1)} \dots
S_\beta^{(9)}\;,\;S_\beta^{(1)}=0\;,\;S_\beta^{(9)}=\infty$, each of
them having the same statistical weight $1/8$, i.e.\ 
\begin{equation}
  \int_{S_\beta^{(i)}}^{S_\beta^{(i+1)}} dS_{\beta}\; P_\beta(S_{\beta}) =
  \frac{1}{8} \;,\quad i = 1 \dots 8 \;,
\end{equation}
and averaged $\hat\varrho_{\alpha\beta}(u_{\alpha};S_\beta)$ in each of
these intervals.

\begin{figure}[!t]
  \begin{center}
    \epsfig{file=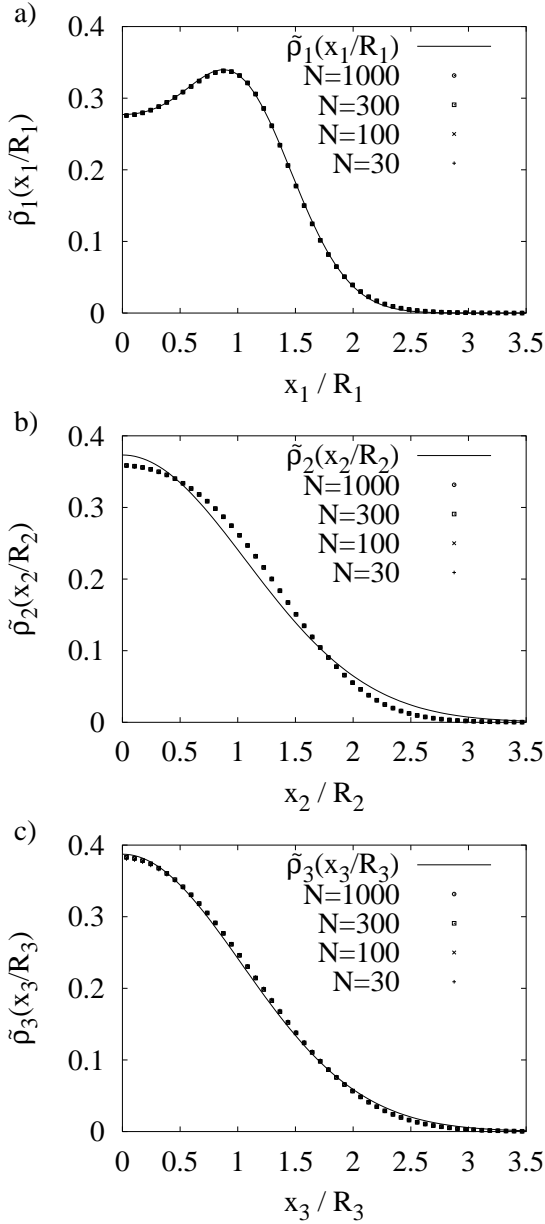,width=0.9\linewidth,angle=0}
  \end{center}
  \caption{Scaled one-dimensional density functions for different
    values of $N$. The straight lines denote the approximate forms
    given in eq.~(\ref{eq:rhoa.heuristic}a-c).}
  \label{fig:density}
\end{figure}
Figure~\ref{fig:conditional.density}a shows the eight resulting
functions for $\hat\varrho_{11}(u_1;S_1,N)$. By contrast to the
self-similarity assumption, the curves differ significantly from each
other. They exhibit an unimodal shape only for the smallest $S_1$,
while for the most likely $S_1$, $\hat\varrho_{11}(u_1;S_1,N)$ has a
bimodal shape that becomes more pronounced for larger $S_1$. A similar
bimodal shape for Gaussian coils has been found for the average
monomer density $\bar\varrho(\textbf{x},N)\!\equiv\!\int d^3S\,
P(\textbf{S})\varrho(\textbf{x};\textbf{S},N)$ by Janszen et.\ al.
\cite{Janszen/et.al.:1996}, and earlier for a model representing
polypropylene by Theodorou and Suter \cite{Theodorou/Suter:1985}.  For
$\hat\varrho_{22}(u_2;S_2)$ shown in
Fig.~\ref{fig:conditional.density}b, the deviations from scaling are
smaller and are appreciable only for the largest $S_2$, while for
$\hat\varrho_{33}(u_3;S_3)$ in Fig.~\ref{fig:conditional.density}c the
deviations are insignificant. We also examined the six other functions
and found that the deviations from scaling are small in all these
cases.

In fact, knowing that the shape of $\bar\varrho(\textbf{x},N)$ in the
$x_{1}$-direction is of bimodal type, while in the $x_{2}$-direction
it is of unimodal type \cite{Janszen/et.al.:1996}, the breakdown of
the self-similarity concept is not that unexpected. If scaling would
hold true, the monomer densities with respect to some fixed
$\textbf{S}$ should correspond to the average monomer density
$\bar\varrho(\textbf{x},N)$. However, if one considers the case of
almost equal values of $S_1$ and $S_2$, i.e.\ $S_1 \gtrapprox S_2$,
the monomer densities along the $x_{1}$- and $x_{2}$-direction should
be the same due to symmetry in contrast to the scaling prediction.

Although the self-similarity assumption is not truely valid, it can be
used as a good working tool for the following reasons. The monomer
densities enter the calculation of $F_{\text{inter}}$ via integrals,
cf.\ eq.~(\ref{eq:Finter}).  For these integrals to be correctly
calculated, it is important that the size of the region, where
$\varrho(\textbf{x},\textbf{S},N)$ is non-neglible, scales properly.
From Fig.~\ref{fig:conditional.density} one can see that this is
indeed the case. Therefore, for our purposes it is sufficient to
approximate the various functions shown in
Fig.~\ref{fig:conditional.density}a-c for different $S_\beta$ by their
average
\begin{equation}
\tilde\varrho_{\alpha}(u_{\alpha})\!=\!
\int_0^\infty d S_\beta\, P_\beta(S_\beta)\,
\hat\varrho_{\alpha\beta}(u_{\alpha};S_\beta,N)\,.
\label{eq:rhoeff}
\end{equation}
We marked these averaged monomer densities by a tilde, since one can
view them as ``effective scaling functions'' representing
$\tilde\varrho(\textbf{u})$ from eq.~(\ref{eq:selfsimilar.density})
(after integrating out the two $u_\mu$-coordinates orthogonal to
$u_{\alpha}$). Note that we suppressed the $N$-argument in the
definition (\ref{eq:rhoeff}), since, as already mentioned, the
dependence on $N$ predicted by eq.~(\ref{eq:selfsimilar.density}) is
valid. This validity can also be inferred directly from
Fig.~\ref{fig:density}, where the data for $N=30,100,300$, and 1000
can not be distinguished.

\begin{table}[!b]
  \caption{Numerical results for the constants defining
      $\tilde \varrho_{\alpha}(u_{\alpha})$ in
      eqs.~(\ref{eq:rhoa.heuristic}a-c, \ref{eq:better.rhoy}).}
  \begin{center}
    \begin{tabular}{|lll|}
      \hline
      $\sigma_1 = 0.48920$ & $c_1 = 0.64528$ & $\bar u_1 = 0.994$ \\
      $\sigma_2 = 1.06892$ &                 &                    \\
      $\sigma_3 = 1.03009$ &                 &                    \\
      \hline
      $\sigma'_2 = 0.748$  &                 & $\bar u_2 = 0.674$ \\
      \hline      
    \end{tabular}
  \end{center}
  \label{tab:rho.constants}
\end{table}
The scaled monomer densities $\tilde\varrho_{\alpha}(u_{\alpha})$ with
$u_\alpha\!=\!x_\alpha/R_\alpha$ shown in Fig.~\ref{fig:density} can
be conveniently approximated by a superposition of Gaussians
\begin{mathletters}
\label{eq:rhoa.heuristic}
\begin{eqnarray}
\tilde\varrho_1(u_1)&=& \frac{1}{\sqrt{2 \pi} \sigma_{1} (2 +
    c_1)} \left[ \exp \left(-\frac{(u_1-\bar u_1)^{2}}{2
        \sigma_{1}^2}\right) + \nonumber \right.\\
&&\hspace*{-0.5cm}\left. \exp \left(-\frac{(u_1+\bar u_1)^{2}}{2
    \sigma_1^{2}} \right) + c_1 \exp \left(-\frac{u_1^{2}}{2 \sigma_1^{2}}
    \right) \right] \label{eq:rhoa.heuristic-a}\\
  \tilde\varrho_2(u_2) &=& \frac{1}{\sqrt{2 \pi} \sigma_2}\,
  \exp\left(-\frac{u_2^{2}}{2 \sigma_{2}^{2}} \right)
\label{eq:rhoa.heuristic-b} \\
  \tilde\varrho_3(u_3) &=& \frac{1}{\sqrt{2 \pi} \sigma_3}\,
  \exp \left(-\frac{u_3^{2}}{2 \sigma_3^{2}} \right)
\label{eq:rhoa.heuristic-c} 
\end{eqnarray}
\end{mathletters}
The parameters $\sigma_\alpha$, $c_1$, and $\bar u_1$ were determined
by least squares fits and are given in Tab.~\ref{tab:rho.constants}.
While $\tilde\varrho_1(u_1)$ and $\tilde\varrho_3(u_3)$ are
quite well approximated by eqs.~(\ref{eq:rhoa.heuristic-a}) and
(\ref{eq:rhoa.heuristic-c}), respectively,
eq.~(\ref{eq:rhoa.heuristic-b}) does not provide such a good
description.  A better account for $\tilde\varrho_2(u_2)$ can be
achieved by adding shifted Gaussian functions as in
(\ref{eq:rhoa.heuristic-a}), and is given explicitly in
appendix~\ref{app:better.rhoy}. However, we decided to deal with
eq.~(\ref{eq:rhoa.heuristic-b}), because the better description
increases the computation time of $F_{\text{inter}}$ by a factor of
four, but does not much improve the overall accuracy of
$\varrho(\textbf{x};\textbf{S})$. We note that the functions
$\tilde\varrho_{\alpha}(u_{\alpha})$ in
eqs.~(\ref{eq:rhoa.heuristic}a-c) should not be confused with the
function $\bar\rho(\textbf{x},N)$ (or its ``components'' after
integrating out two coordinates) that has been considered earlier by
Janszen et al.~\cite{Janszen/et.al.:1996}.

In order to specify the multivariate scaled density
$\tilde\varrho(\textbf{u})$ we employ, as for $P(\textbf{S})$ before,
a separation ansatz. This yields for
$\varrho(\textbf{x};\textbf{S},N)$ from
eq.~(\ref{eq:selfsimilar.density}) the final result
\begin{equation}
\varrho(\textbf{x};\textbf{S},N)=
(N\!+\!1)\prod_{\alpha}\frac{\tilde\varrho(x_\alpha/R_\alpha)}{R_\alpha}\,,
\label{eq:varrho}
\end{equation}
where $\tilde\varrho(x_\alpha/R_\alpha)$ are taken from
eqs.~(\ref{eq:rhoa.heuristic}a-c). The separation ansatz may be used
also without invoking the self-similarity assumption in order to
reduce the complexity of the multivariate density
$\varrho(\textbf{x};\textbf{S},N)$.

\begin{table}[!b]
  \caption{Test of the ansatz eq.(\ref{eq:varrho}) with
      correlation coefficients defined in eq.~(\ref{eq:gamma-v}).}
  \begin{center}
    \begin{tabular}{|cl|}
      \hline 
      $(\langle v_{1} v_{2} \rangle - \langle v_{1} \rangle \langle
      v_{2} \rangle) / \langle v_{1} v_{2} \rangle$ & -0.048 \\ 
      $(\langle v_{2} v_{3} \rangle - \langle v_{2} \rangle \langle
      v_{3} \rangle) / \langle v_{2} v_{3} \rangle$ & -0.027 \\ 
      $(\langle v_{3} v_{1} \rangle - \langle v_{3} \rangle \langle
      v_{1} \rangle) / \langle v_{3} v_{1} \rangle$ & -0.018 \\ 
      \hline
    \end{tabular}
  \end{center}
  \label{tab:rho.correlation.coefficients}
\end{table}
To test the validity of the separation ansatz we define for each
Gaussian chain by $v_\alpha\!=\!(N\!+\!1)^{-1}\!\sum_{k=0}^N
|x_{\alpha}^{(k)}|/R_\alpha$ the (scaled) mean moduli of the monomer
coordinates in the principal axis system, and check whether the
correlation coefficients
\begin{equation}
\gamma_{\alpha\beta}^{(v)} \equiv\frac{\langle v_{\alpha}v_{\beta}
  \rangle - \langle v_{\alpha} \rangle \langle v_{\beta} \rangle} 
  {\langle v_{\alpha}v_{\beta} \rangle}\,,\hspace*{0.3cm}\alpha\ne\beta\,,
\label{eq:gamma-v}
\end{equation}
are much smaller than one. Note that, because of the symmetry
$\tilde\varrho_\alpha(u_\alpha)=\tilde\varrho_\alpha(-u_\alpha)$, we
considered the moments of the moduli in (\ref{eq:gamma-v}). As shown
in Tab.~\ref{tab:rho.correlation.coefficients}, we find
$|\gamma_{\alpha\beta}^{(v)}|$ to be less than 5\% in the simulations.
\begin{figure}
  \begin{center}
    \epsfig{file=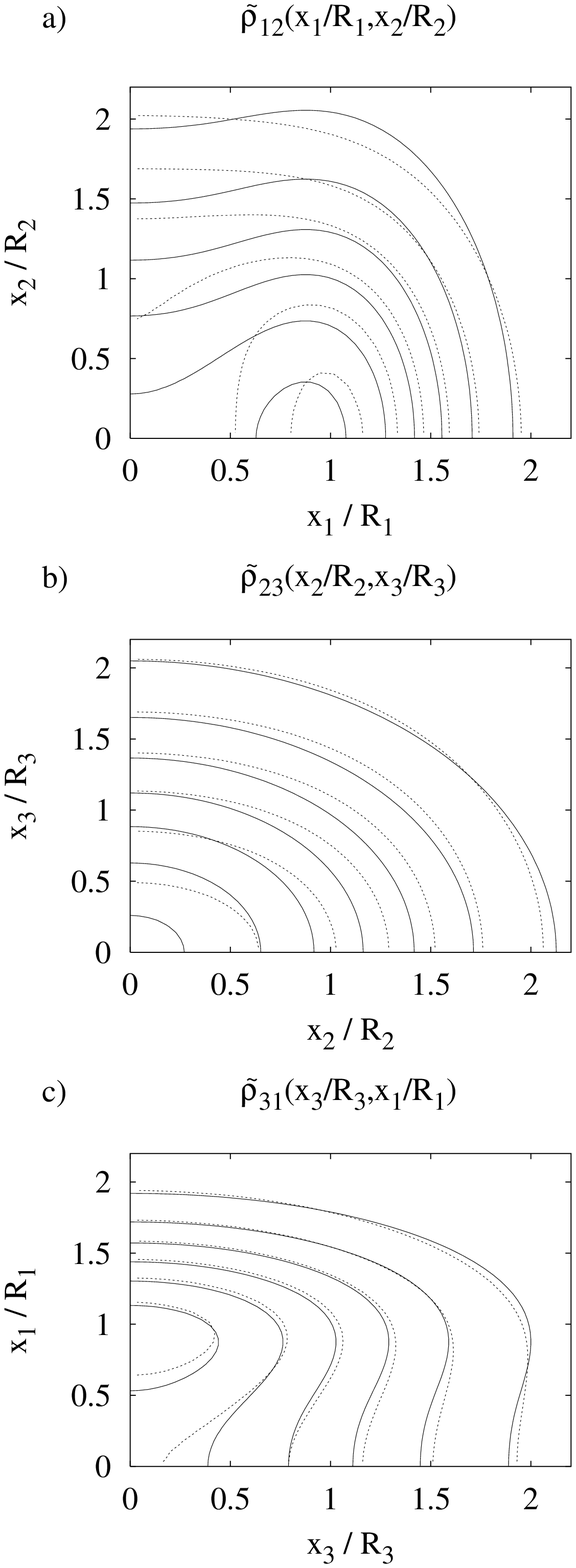,width=0.75\linewidth,angle=0}
  \end{center}
  \caption{Contour plots showing the comparison of
    $\tilde \varrho_{\alpha}(x_{\alpha}/R_{\alpha}) \tilde
    \varrho_{\beta}(x_{\beta}/R_{\beta})$ (straight lines) with
    numerical data of $\tilde
    \varrho_{\alpha\beta}(x_{\alpha}/R_{\alpha}, x_{\beta}/R_{\beta})$
    (dashed lines). The isolines are drawn at function values, which
    are integer multiples of $0.02$.}
  \label{fig:rho.comparison}
\end{figure}
Also, when comparing the two-variate densities
\begin{equation}
\tilde\varrho_{\alpha\beta}(u_{\alpha},u_{\beta}) =
  \int_{-\infty}^{\infty} du_{\gamma}\, \tilde\varrho(\textbf{u}) \;,\quad
  \gamma \neq \alpha, \beta 
\end{equation}
with the two-fold products $\tilde\varrho_{\alpha}(u_{\alpha})
\tilde\varrho_{\beta}(u_{\beta})$ in Fig.~\ref{fig:rho.comparison},
a satisfactory agreement is obtained.

\begin{figure}[!t]
  \begin{center}
    \epsfig{file=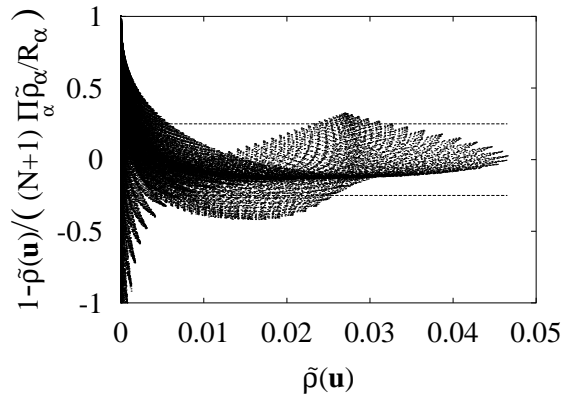,width=0.9\linewidth,angle=0}
  \end{center}
  \caption{The plot shows the relative error of the approximate
    function defined in
    eqs.~(\ref{eq:rhoa.heuristic},\ref{eq:varrho}) when compared to
    3D numerical data of $\tilde \varrho(\textbf{u})$ for chains of
    length $N=1000$, versus the function value of $\tilde
    \varrho(\textbf{u})$. For large function values, that give the
    main contribution to $F_{\text{inter}}$, the relative error is
    quite small. The lines denote a 25\% error interval.}
  \label{fig:rho.relative.error}
\end{figure}
As an ultimate test of the separation ansatz we sampled the full
three-variate scaled monomer density $\tilde\varrho(\textbf{u})$ in
the simulations. To this end, we subdivided the
$\textbf{u}=\textbf{x}/\textbf{R}$ space into small boxes of length
$\Delta u_\alpha\!=\!0.07$ and determined the mean number of monomers
in each box by averaging over Gaussian chains irrespective of their
$\textbf{S}$. The relative error between the separation ansatz
$\prod_{\alpha}\tilde\varrho_\alpha(u_\alpha)$ with $\tilde
\varrho_\alpha(u_\alpha)$ from eqs.~(\ref{eq:rhoa.heuristic}a-c) and
the simulated $\tilde\varrho(\textbf{u})$ is plotted against
$\tilde\varrho(\textbf{u})$ in Fig.~\ref{fig:rho.relative.error} for
all $\textbf{u}$. We see that the relative error is confined to a
narrow band of maximal 25\% discrepancy except for very small
$\tilde\varrho(\textbf{u})$. These small $\tilde\varrho(\textbf{u})$,
however, are not important in the calculation of the overlap
integrals in eq.~(\ref{eq:Finter}). In principle, it would be
desirable to do the same analysis also with $P(\textbf{S})$ but since
one Gaussian chain generates only one realization of $\textbf{S}$, it
is very hard to obtain reasonable statistics in this case.

The monomer density $\varrho(\textbf{x},\textbf{S},N)$ specified in
eq.~(\ref{eq:varrho}) allows us to calculate the overlap integrals and
thus $F_{\text{inter}}$ from (\ref{eq:Finter}) analytically.  This is
explained in detail in Appendix~\ref{app:analytic.integration}.

\subsection{Self-interactions}
\label{subsec:selfinter}

Gaussian chains are by definition non-self-interacting in contrast to
real polymer chains. In order to obtain a physical model, we therefore
include a self-interaction term in eq.~(\ref{eq:Fintersum}), which
amounts to an intermolecular part of the form
\begin{equation}
F_{\text{inter}} = \frac{1}{2} \sum_{i=1}^{M} \sum_{j \neq i}
  F_{\text{inter}}^{(ij)}+
\frac{1}{2} \sum_{i=1}^{M}F_{\text{inter}}^{(ii)}
\end{equation}
with $F_{\text{inter}}^{(ij)}$ given in eq.~(\ref{eq:Finter}). The
factors (1/2) in front of both the double and simple sum in
(\ref{eq:Fintersum}) arise to avoid double counting of monomer
interactions.

In principle, it would be more appropriate to include the
self-interaction term into the intramolecular part in
eq.~(\ref{eq:Fintra}).  In fact, when referring to some kind of
realistic model for single polymer chains, as e.g.\ the bead-spring
model considered by Murat and Kremer \cite{Murat/Kremer:1998} this
self-interaction part naturally is entailed in $F_{\text{intra}}$. In
our case, however, interactions between monomers belonging to the same
or different chains are treated {\it on the same footing} and we thus
decided to include the self-interaction part into
eq.~(\ref{eq:Fintersum}). The GEM thus specified can be regarded as a
Flory type model for polymer systems.

\subsection{Remarks on the algorithm}
\label{sec:algorithm}

In this section we shortly describe some details of the algorithm used
throughout the rest of this work. As the Monte-Carlo method is well
established \cite{Binder/ed.:1995}, we just comment on some
specialties in this case.

The main challenge for the setup of an optimal algorithm is to
calculate the interaction term $F_{\text{inter}}$ efficiently, even
when the calculation can be done analytically. As one polymer in a
melt overlaps with many other ones, it is crucial to calculate only
those interaction terms, which contribute significantly to
$F_{\text{inter}}$. Therefore, the monomer density has to be cutoff at
a certain length and we choose this cutoff to be
$x_{\alpha,\text{cut}}/R_{\alpha}=2.2$, cf.\ Fig.~\ref{fig:density}.
The neighbors of a polymer are defined as those polymers, which
interpenetrate each other within their cutoff shape. Because the
ellipsoids are strongly aspherical, a simple test of the distance
between the centers of mass is not efficient enough.

In order to determine whether two ellipsoids are neighbors, we
introduced a grid of cells, with a spatial discretization length given
approximately by the smallest significant scale, i.e.\ $\langle
R_3\rangle$. To every cell in the grid is assigned a list of polymers,
that overlap with the cell, and conversely, to every polymer is
assigned a list of cells, which it covers.  Determining the neighbors
$j$ of one polymer $i$ then is simple: One goes through the list of
cells assigned to polymer $i$, determines all polymers $j$ that are
associated with each member of this list and finally sorts out
multiple occurrences of $j$. For each polymer we keep a list entailing
the neighbors with the respective interaction energies, and the total
value of the interaction energy with the neighbors. In this way the
updating of the lists after each Monte-Carlo move becomes simple.
Since the calculation of $F_{\text{inter}}$ consumes the main part of
the CPU time, we parallelized this part, thus speeding up the
simulation by a factor of $1.5$ on a two-processor machine.

With the method described here, a fast calculation of $F_{\text{inter}}$ is
possible. However it should be noted, that the efficiency depends on the
possibility to integrate the product of two densities analytically. The time
for calculation depends essentially both on the overall monomer concentration
$c$ and on $N$ (or on $\bar R_G$), because these quantities determine the
average number of neighbors of one particle.

For performing one elementary step of the Monte-Carlo simulation,
where we try to change either the orientation, the center of mass
position or the eigenvalues $\textbf{S}$ of one randomly chosen
ellipsoid as explained in Sec.~\ref{sec:model}, we still have to
specify $\Delta r_{\text{max}}$ and $\Delta S_{\text{max}}$. We
choose
\begin{equation}
  \label{eq:Delta.r}
\Delta r_{\text{max}}=K_r \bar R_G\,,\hspace*{0.5cm}
\Delta S_{\text{max}}=K_s \langle S_\alpha \rangle
\end{equation}
with $K_r\!=\!0.25$, $K_s\!=\!0.5$ and the $\langle S_{\alpha}
\rangle$ given in Tab.~\ref{tab:Sa.moments}.  A trial for a change
from an initial state {\sl i} to a final state {\sl f} is accepted
with a probability $p_{\sl if}$ given by the Glauber rule, i.e.\ 
$p_{\sl if}\!=\!(1/2)(1-\tanh[(F_{\sl f}-F_{\sl i})/2k_{\rm B}T])$,
where $F_{\sl i}$ and $F_{\sl f}$ are the free energies of the initial
and final state, respectively. After each trial the time $t$ is
incremented by $\tau_0/3M$, where $\tau_0$ is a characteristic time
scale (1~MCS) that has to be adapted to the particular physical
situation under consideration (see the discussion in Sec.~\ref{sec:melt}
below).

Unless noted otherwise, the initial conditions are as follows: The
center of mass positions of the ellipsoids are randomly distributed
within the simulation box. The orientations of the ellipsoids are
random, and the eigenvalues $\textbf{S}_i$ are drawn from the
distribution $P(\textbf{S},N)$ for the non-interacting Gaussian chains
given in eq.~(\ref{eq:Sa.heuristic}).  The $S_\alpha$ are sorted
afterwards according to rule (\ref{eq:Sa.condition}). Equilibrium
quantities are determined after proper thermalization.

\vspace*{0.3cm} To summarize this section, eqs.~(\ref{eq:prod.PS}) and
(\ref{eq:varrho}) together with (\ref{eq:rhoa.heuristic}) define the
input quantities for the GEM, and one has to take care of including
self-interactions as explained in Sec.~\ref{subsec:selfinter}. The
explicit form of the respective formulae allows for an easy
reproducibility of the results to be discussed in the following
sections. Despite approximate, they capture the essential features
found for Gaussian chains: The normal random walk scaling with $N$,
the strong asphericity of the chains, the bimodal form of the monomer
density and its rapid decay beyond cutoff lengths that scale linearly
with $R_\alpha$. We note in passing that we performed also simulations
for self-avoiding chains and found similar features except, of course,
for the scaling with $N$.  While it is possible to provide more
accurate formulae for the input quantities by introducing more
parameters, we are convinced that this is of minor importance for the
study of structural properties and kinetic phenomena on
semi-macroscopic length scales.


%% file: melt.cm.tex
\section{Homogeneous Systems}
\label{sec:melt}
The aim of the following section is to show that the GEM reproduces
the most important semi-macroscopic properties of dilute and dense
polymeric systems: The scaling properties of the radius of gyration
$\bar R_G$ upon $N$ and $c$, the scaling behavior of the distribution
$P_R(R_{G}^{2},N)$ in a dense homogenous system of interacting
ellipsoids, and the existence and scaling of the correlation hole. We
partly follow the analysis proposed by Murat and Kremer
\cite{Murat/Kremer:1998}.

\begin{figure}[!t]
  \begin{center}
    \epsfig{file=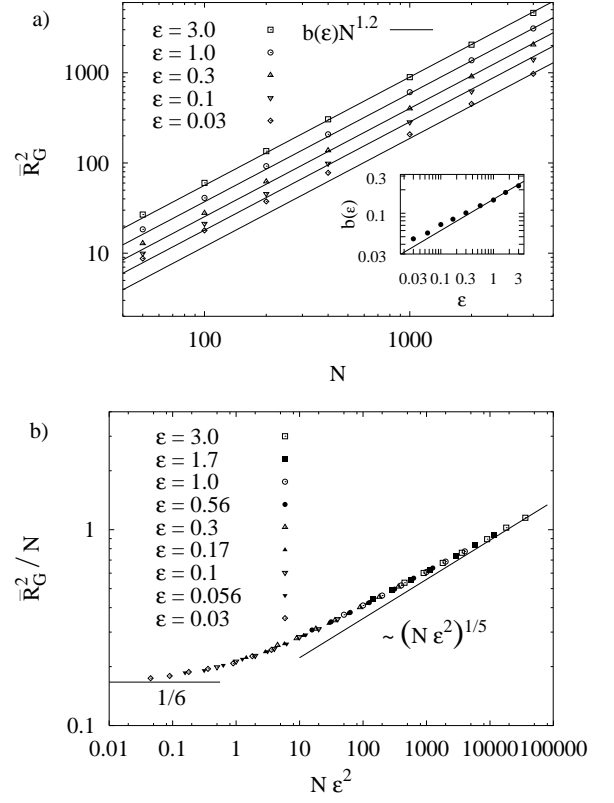,width=0.9\linewidth,angle=0}
  \end{center}
  \caption{a) Mean squared gyration radius of free ($c \to 0$), but
    self-interacting ellipsoids as a function of $N$, displayed for
    various values of the interaction strength $\epsilon$. The Flory
    result $\langle R_{G}^{2} \rangle \propto N^{6/5}$ is shown for
    comparison. The inset shows the dependence on $\epsilon$. The line
    indicates a power law $b(\epsilon) \propto \epsilon^{2/5}$.  b)
    Crossover scaling from Gauss to Flory type behavior.}
  \label{fig:RG2.selfinteracting.free}
\end{figure}
As mentioned above, the free energy as given in
eqs.~(\ref{eq:F}--\ref{eq:Fintersum},\ref{eq:Finter}) is some kind of
Flory-like description of polymers. Accordingly, the gyration radius
of a free ellipsoid, that only interacts with itself but not with
other ellipsoids, scales as $R_{G}^{2}\sim N^{6/5}$
\cite{deGennes:1979}, see Fig.~\ref{fig:RG2.selfinteracting.free}a.

In order to obtain the dependence on $\epsilon$ also, let us redo some
Flory type calculation, where for simplicity we consider soft spheres
instead of soft ellipsoids. Within this approximation the monomer
density from eq.~(\ref{eq:varrho}) would become
\begin{equation}
  \label{eq:density.radial}
  \varrho_r(r;R_G)=\frac{(N\!+\!1)}{R_G^3}\tilde\varrho_r
  \left(\frac{r}{R_G}\right)
\end{equation}
with
$\tilde\varrho_r(u_r)\!=\!\exp(-u_r^2/2\sigma_r^2)/(2\pi\sigma_r^2)^{3/2}$,
$\sigma_r^2\!=\!1/3$. Such a simplified form of the monomer density
has been discussed already in \cite{Yamakawa:1971}. Substituting
$P_{R}(R_G^2)$ from eqs.~(\ref{pr-eq},\ref{fR(x)-eq}) into the free energy
functional $F_{\text{sph}}^{(ii)}(R_{G}^{2})\!=\!-k_{\rm B}T\ln
P_{R}(R_G^2)\!+\!4\pi\hat\epsilon\int_0^\infty\! dr\, r^2
\varrho_r^2(r;R_G)$ we obtain
\begin{eqnarray}
  \label{eq:Fii.self}
  \frac{F_{\text{sph}}^{(ii)}(R_{G}^{2})}{k_{B}T}&=&
  -\ln\left(\frac{1}{2\,N K_{0}(2\,d_{R})}\right)
  -\ln\left(\frac{N}{R_{G}^{2}}\right)\nonumber\\[0.1cm]
&&\hspace*{-1.5cm}  
+\frac{R_{G}^{2}}{a_{R}N} + d_{R}^{2}\frac{a_{R}N}{R_{G}^{2}} +
  \epsilon\frac{(N\!+\!1)^2}{32}
  \left(\frac{3}{\pi\,R_{G}^{2}}\right)^{3/2} \,.
\end{eqnarray}
The most probable value $R_{G,0}^2$ results from
minimizing with respect to $R_G^2$,
\begin{equation}
  \label{eq:condition.free}
  \frac{1}{R_{G,0}^{2}} + \frac{1}{a_{R}N} - d_{R}^{2}
  \frac{a_{R}N}{R_{G,0}^{4}}-  
  \epsilon\frac{3(N\!+\!1)^{2}}{64\,R_{G,0}^{5}}
  \left(\frac{3}{\pi}\right)^{3/2} \hspace*{-0.2cm}= 0\,. 
\end{equation}
This equation has the asymptotic solution
\begin{equation}
R_{G,0}^{2}\sim
\left(\frac{3}{\pi}\right)^{3/5}
\left(\frac{3}{64}\;\epsilon\right)^{2/5} N^{6/5}\,.
\label{rg0-eq}
\end{equation}
Since $\bar R_G^2\!\sim\!R_{G,0}^{2}$, we expect $\bar
R_G^2\!\sim\!\epsilon^{2/5}$ in reasonable agreement with the
numerical results for free, self-interacting ellipsoids, see
Fig.~\ref{fig:RG2.selfinteracting.free}a.

\begin{figure}[!b]
  \begin{center}
    \epsfig{file=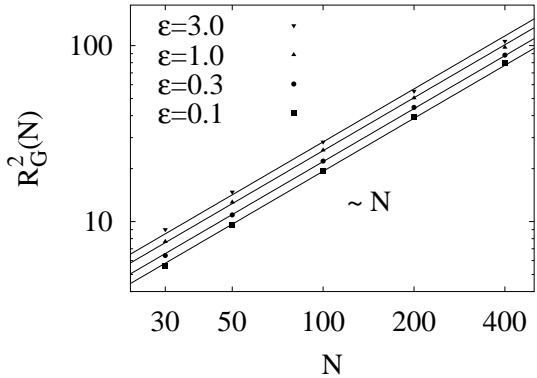,width=0.9\linewidth,angle=0}
  \end{center}
  \caption{Squared gyration radius $R_{G}^{2}$ as a function of $N$ in
    a dense ($c = 0.85$) system of $M=1000 \to 4000$ ellipsoids
    (double-logarithmic representation). For various values of
    constant $\epsilon$ a linear behavior is observed.}
  \label{fig:RG2.selfinteracting.melt}
\end{figure}
\begin{figure}[!t]
  \begin{center}
    \epsfig{file=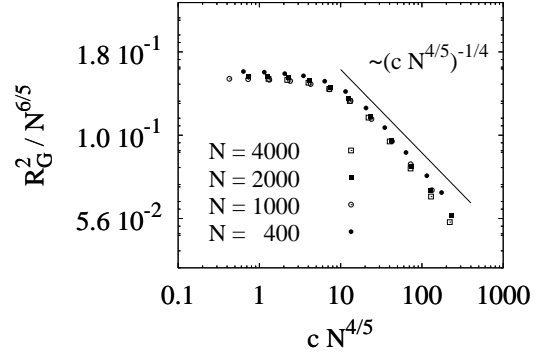,width=1.0\linewidth,angle=0}
  \end{center}
  \caption{Test of the scaling relation between the squared radius of
    gyration and the monomer concentration for various values of $N$
    and constant $\epsilon = 1.0$ (double-logarithmic representation).
    For small concentrations the chains are almost free and
    $R_{G}^{2}$ saturates, corresponding to Flory type behavior. In
    the region of high concentration the expected scaling relation is
    almost fulfilled.}
  \label{fig:RG2.Nb.concentration}
\end{figure}
For small $\epsilon$, one should recover the Gaussian behavior, i.e.\ 
we expect $\bar R_G^2\!\sim\!N$ for
$\epsilon\!\ll\!\epsilon_\star(N)$. On the other hand, the scaling
(\ref{rg0-eq}) must hold true for $\epsilon\!\gg\!\epsilon_\star(N)$.
Thus, from continuity, we find $\epsilon_\star(N)\!\sim\!N^{-1/2}$.
The corresponding scaling form $\bar
R_G^2(N,\epsilon)\!=\!Nf(\epsilon^{2} N)$ with $f(u)\!\sim\!{\rm
  const.}$ for $u\to0$ and $f(u)\!\sim\!u^{1/5}$ for $u\to\infty$ is
verified in Fig.~\ref{fig:RG2.selfinteracting.free}b.

We now discuss systems composed of many interacting ellipsoids. As is
well known from self-consistent field arguments \cite{deGennes:1979},
the Flory type scaling (\ref{rg0-eq}) changes to a normal random walk
scaling if the monomer concentration $c$ exceeds the overlap
concentration $c_\star$. That this is indeed the case is shown in
Fig.~\ref{fig:RG2.selfinteracting.melt}.  Approaching $c_\star$ from
below we can write $(c_\star/N)^{-1/3}\!\sim\!\bar
R_G(N,\epsilon)\!\sim\!  \epsilon^{1/5}N^{3/5}$ (for
$\epsilon\!\gg\!N^{-1/2}$), i.e.\ 
$c_\star\!\sim\!\epsilon^{-3/5}N^{-4/5}$, so that more generally we
expect the scaling form \cite{deGennes:1979}
\begin{equation}
\bar R_G^2(N,\epsilon,c)=\epsilon^{2/5}N^{6/5}f(c\,\epsilon^{3/5} N^{4/5})
\hspace*{0.3cm}
\label{eq:rgscal}
\end{equation}
with $f(\lambda)\!\to\!{\rm const.}$ for $\lambda\!\ll\!1$ and
$f(\lambda)\!\to\!\lambda^{-1/4}$ for $\lambda\!\gg\!1$ to describe
the crossover from dilute to dense systems. Accordingly, $\bar
R_G^2\!\sim\!\epsilon^{1/4}c^{-1/4}N$ for $c\,\epsilon^{3/5}
N^{4/5}\!\gg\!1$. To demonstrate the validity of the overall scaling
predicted by eq.~(\ref{eq:rgscal}), we show in
Fig.~\ref{fig:RG2.Nb.concentration} as an example $\bar R_G^2/N^{6/5}$
as a function of $cN^{4/5}$ for a fixed value $\epsilon\!=\!1.0$. As
can be seen from the figure, the data for different $c$ and $N$ all
collapse onto a common master curve. We note that the scaling
(\ref{eq:rgscal}) is followed only for $\epsilon\!\gg\!N^{-1/2}$. For
$\epsilon\!\ll\!N^{-1/2}$ by contrast, the ellipsoids would always
exhibit Gauss type behavior.

\begin{figure}[!t]
  \begin{center}
    \epsfig{file=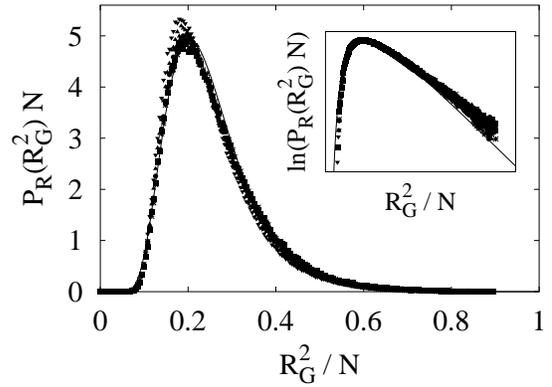,width=0.9\linewidth,angle=0}
  \end{center}
  \caption{Scaling behavior of the distribution function
    $P(R_{G}^{2})$ in a melt system for $N =$ 30 ($\ast$), 50
    ($\blacksquare$), 100 ($\bullet$), 200 ($\blacktriangle$) and 400
    ($\blacktriangledown$). The straight line refers to a function of
    form eq.~(\ref{fR(x)-eq}), but with changed parameters $a_{R} =
    0.0665$ and $d_{R} = 3.52$. The inset shows the same data in
    semi-logarithmic representation. The external parameters are
    $c=0.85$ and $\epsilon = 1.0$.}
  \label{fig:P.RG2.s.skalen.melt}
\end{figure}

We further show in Fig.~\ref{fig:P.RG2.s.skalen.melt} that the
distribution function $P_R(R_{G}^{2},N)$ for dense systems
($c\!\gg\!c_\star$) obeys the same type of scaling as for
non-interacting Gaussian chains, eq.~(\ref{pr-eq}),
$P_R(R_{G}^{2},N)\!\sim\!N^{-1}\tilde P_R(R_G^2/N)$.  In particular,
the ansatz (\ref{fR(x)-eq}) can be used equally well for $\tilde
P_R(u)$ up to a change of the parameters $a_{R}$, and $d_{R}$
(see Tab.~\ref{tab:probability.RG2.constants} and the values given in
the caption of Fig.~\ref{fig:P.RG2.s.skalen.melt}).

\begin{figure}[!b]
  \begin{center}
    \epsfig{file=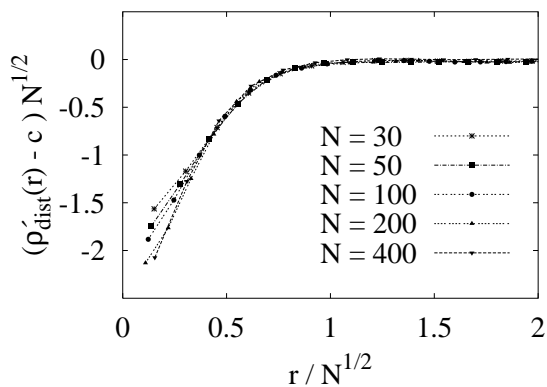,width=0.9\linewidth,angle=0}
  \end{center}
  \caption{Scaling behavior of the correlation hole. The monomer
    concentration is $c=0.85$, the interaction parameter $\epsilon =
    1.0$.}
  \label{fig:correlation.hole}
\end{figure}
\begin{figure}[!t]
  \begin{center}
    \epsfig{file=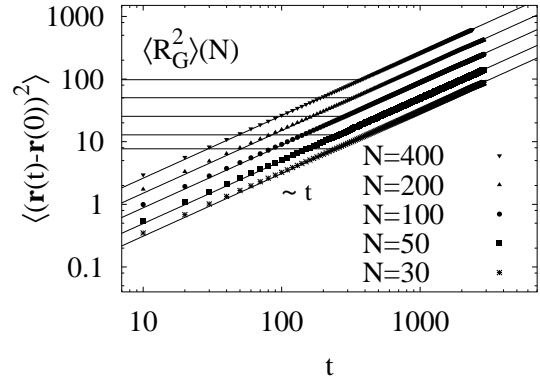,width=0.9\linewidth,angle=0}
  \end{center}
  \caption{Mean square displacement $\langle
    (\textbf{r}(t)-\textbf{r}(0))^{2} \rangle$ as a function of time
    for various values of $N$ in a dense system ($c=0.85$,
    $\epsilon=1.0$). The lines indicate a linear growth at late times.
    Shown for comparison is the mean squared radius of gyration
    $\langle R_{G}^{2} \rangle$ in the melt for the respective $N$.}
  \label{fig:mean.square.displacement}
\end{figure}
To complete our study of the static properties of dense homogeneous
systems, we discuss the properties of the so-called correlation hole
\cite{deGennes:1979}. Therefore we determine the mean monomer density
of all ellipsoids except the $i$'th one as a function of the distance
from the center of mass of ellipsoid $i$,
\begin{equation}
\varrho'_{\rm  dist}(r,N)=\Bigl\langle\sum_{j\ne i}
\varrho'_j(\textbf{r}-\textbf{r}_i,N)\Bigr\rangle\,.
\label{eq:rhodist}
\end{equation}
After performing the average in (\ref{eq:rhodist}), this mean monomer
density of ``distinct ellipsoids'' is equal for all $i$ and depends on
$r=|\textbf{r}|$ only. For large $r\!\gtrsim\!\bar R_G$,
$\varrho'_{\rm dist}(r)\!\to\!c$, while for small $r\!\ll\!\bar R_G$,
$\varrho'_{\rm dist}(r)$ must be smaller than $c$ due to the fact that
ellipsoid $i$ has been excluded from the sum in
eq.~(\ref{eq:rhodist}).  Since $4\pi\int_0^\infty
dr\,r^2\,[\varrho'_{\rm dist}(r,N)-c]=-(N\!+\!1)$, we expect a scaling
$[\varrho'_{\rm dist}(r,N)\!-\!c]\!\sim\!N^{-1/2}f(rN^{-1/2})$ for the
correlation hole, in reasonable agreement with the simulated results
shown in Fig.~\ref{fig:correlation.hole}.

We conclude this section with some remarks on the dynamical behavior
of the system.  As can be seen from
Fig.~\ref{fig:mean.square.displacement}, the time-dependent mean
square displacement
$\langle[\textbf{r}_i(t)-\textbf{r}_i(0)]^2\rangle$ of an ellipsoid
exhibits normal diffusive behavior for times $t\!>\!\tau_D$, where
$\tau_D$ is the disengagement time, i.e.\ the average time an
ellipsoid needs to diffuse over a distance $\bar R_G$. The short time
regime $t\!<\!\tau_D$ is not of interest here, since we do not intend
to capture the complicated dynamics of polymer systems on these time
scales. In particular, the diffusion coefficient $D=\lim_{t\to\infty}
\langle[\textbf{r}_i(t)-\textbf{r}_i(0)]^2\rangle/6t$ will usually not
exhibit the desired scaling with $N$, e.g.\ $D\!\sim\!N^{-1}$ for
Rouse chains or $D\!\sim\!N^{-2}$ for entangled chains.

In fact, because $D$ depends on $\Delta r_{\rm max}$, and $\Delta
r_{\rm max}$ is chosen to be some fraction $K_r$ of $\bar
R_G\!\sim\!N^{1/2}$ (see Sec.~\ref{sec:algorithm}), the variation of
$D$ with $N$ is influenced by the simulation procedure itself, i.e.\ 
by the value of $K_r$. For the choice $K_r=0.25$ used in our
simulations, $D$ increases with $N$ (see
Fig.~\ref{fig:mean.square.displacement}). Conceptually, this
is not a crucial problem. Since the elementary time scale $\tau_0$ in
the Monte Carlo procedure (see Sec.~\ref{sec:algorithm}) is arbitrary,
we can always adjust $\tau_0\!=\!\tau_0(N)$ in order to reproduce the
desired $N$ dependence.

Let us finally note that apart from the characterstic time scale
$\tau_D$ for translational motion, there are two other characterstic
time scales in the model, which refer to the change of shape and
orientation of the ellipsoids.  These can be identified by the decay
of e.g.\ the correlation functions $\langle R_{G}^{2}(t)
R_{G}^{2}(0)\rangle$ (shape) and $\langle
\textbf{e}_1(t)\cdot\textbf{e}_1(0) \rangle$ (orientation), where
$\textbf{e}_1$ is a unit vector pointing in the direction of the first
principal axis (normalized first eigenvector of $S_{\alpha\beta}$). An
adjustment of these time scales to real situations is again possible
by tuning $K_S$ properly, or by restricting the rotational movements
of the ellipsoids.  For our choice of parameters ($K_S\!=\!0.5$ and
free rotation), $\tau_D$ is the largest characteristic time scale.
Accordingly, for the time scales $t\!\gtrsim\!\tau_D$ of interest
here, we have not to bother about shape-sensitive or orientational
correlations.


%% file: mixture.cm.tex
\section{Binary mixtures}
\label{sec:mixture}

In this section we present the results for a binary mixture of
polymers. We denote the fraction of chains of type $A$ with
$f_A=M_A/M$, and the fraction of chains of the other type $B$ with
$f_B=M_B/M$. The interaction parameter is always $\epsilon=1.0$ and we
consider dense systems ($c=0.85$).

\subsection{Coexistence curve}

First we determine the coexistence curve.  The simulations are
conducted for a semi-grandcanonical ensemble \cite{Murat/Kremer:1998},
i.e.\ we keep the total particle number $M$ fixed and allowed the
fractions $f_A$ and $f_B$ to fluctuate. To this end we allow
particles to change their identity in the Monte-Carlo simulation in
addition to the other possible moves. The order parameter $\Phi$ is
defined by $\Phi\!\equiv\!|f_A-f_B|$. For $T\!<\!T_c$ the distribution
of the order parameter $D(\Phi)$ has a maximum at a finite value
$\Phi_{\rm max}\!>\!0$, while for $T\!>\!T_c$, $\Phi_{\rm max}\!=\!0$.

\begin{figure}[!b]
  \begin{center}
    \epsfig{file=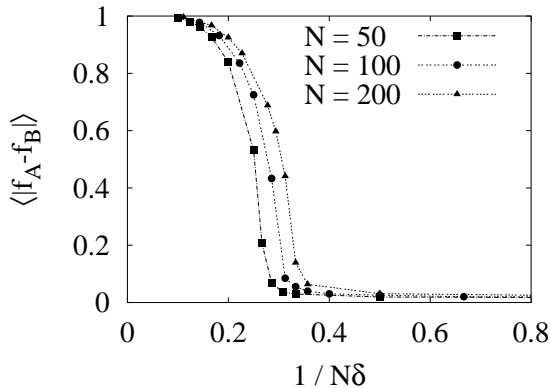,width=0.9\linewidth,angle=0}
  \end{center}
  \caption{The order parameter $\langle|f_{A}-f_{B}|\rangle$ plotted
    versus $1/N\delta\!=\!k_{\rm B}T/N(\hat\epsilon_{AB} -
    \hat\epsilon)$ for chains of length $N=50, 100$ and $200$. The
    result indicates, that the coexistence curve scales almost as
    $1/N$.}
  \label{fig:phasediagramms}
\end{figure}

A convenient way to identify the phase transition in the simulation
is to determine the average $\langle\Phi\rangle$. For $T\!\ll\!T_c$,
$D(\Phi)$ becomes sharply peaked and
$\langle\Phi\rangle\!\simeq\!\Phi_{\rm max}$. For $T\!>\!T_c$,
$\langle\Phi\rangle$ is measure of the width of $D(\Phi)$ rather than
an estimate of $\Phi_{\rm max}\!=\!0$. However, for $T\!\gg\!T_c$ the
width becomes very small and $\langle\Phi\rangle$ also.  Accordingly,
the critical temperature can be estimated from the region, where
$\langle\Phi\rangle$ goes from large finite values to zero, see
Fig.~\ref{fig:phasediagramms}.

We investigated systems of $M=4000, 3000$ and $2000$ ellipsoids for
chain lengths $N=50, 100$ and $200$ respectively, carrying out runs
over 4000 MCS. In Fig.~\ref{fig:phasediagramms} we show
$\langle\Phi\rangle\!=\!\langle|f_A-f_B|\rangle$ as a function of
$1/N\delta\!=\!k_{\rm B}T/N(\hat\epsilon_{AB}-\hat\epsilon)$. The
scaling of the coexistence curve with $N$ is not perfect since $k_{\rm
  B}T_c/\delta=A\,N\!+\!B$ \cite{Binder:1995,Deutsch/Binder:1993}. A
similar behavior of the phase diagram in dependence on $N$ has been
found in simulations with explicit chains and in experiments
\cite{Gehlsen/et.al.:1992}.

\subsection{Chain dimensions in a blend}

\begin{figure}[!t]
  \begin{center}
    \epsfig{file=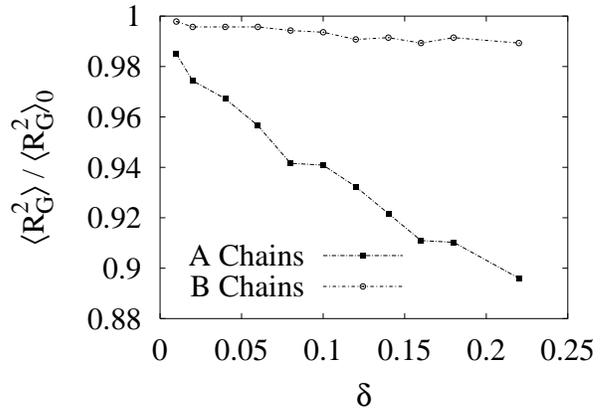,width=0.95\linewidth,angle=0}
  \end{center}
  \caption{Squared radius of gyration of minority
    $(f_{A}=0.1)$ and majority $(f_{B}=0.9)$ type chains as a function
    of the excess interaction parameter $\delta$. The simulation was
    run for $M=4000$ particles of length $N=50$.}
  \label{fig:minority.shrinking}
\end{figure}
An interesting physical effect in polymer blends is that the chains of
the minority phase reduce their size to minimize the repulsive
interaction with those of the majority phase
\cite{Binder:1995,Sariban/Binder:1988}. In
Fig.~\ref{fig:minority.shrinking} we present the results for the
average squared gyration radius of the two different components $A$
and $B$, divided by the pure melt value $\langle R_{G}^{2}
\rangle_{0}$ for $\delta=0$. The fraction of $B$-particles was kept
fixed at $f_{B}=0.9$. As can be seen from the figure, the particles
in the minority shrink much more strongly, than those in the majority.
A detailed discussion of the interplay between thermodynamics and
chain conformations can be found in \cite{Mueller:1998}.

\subsection{Spinodal decomposition}

When a polymer mixture is quenched rapidly into the mechanically
unstable part of the phase diagram, demixing occurs via spontaneously
growing concentration fluctuations of long wavelengths
\cite{Bray:1994}. After some initial time regime, domains of the two
coexisting phases form, and with increasing time these domains are
expected to coarsen according to the Lifshitz-Slyozov growth law
$L_D(t)\!\sim\!t^{1/3}$, where $L_D$ denotes the average domain size.
\begin{figure}[!t]
  \begin{center}
    \epsfig{file=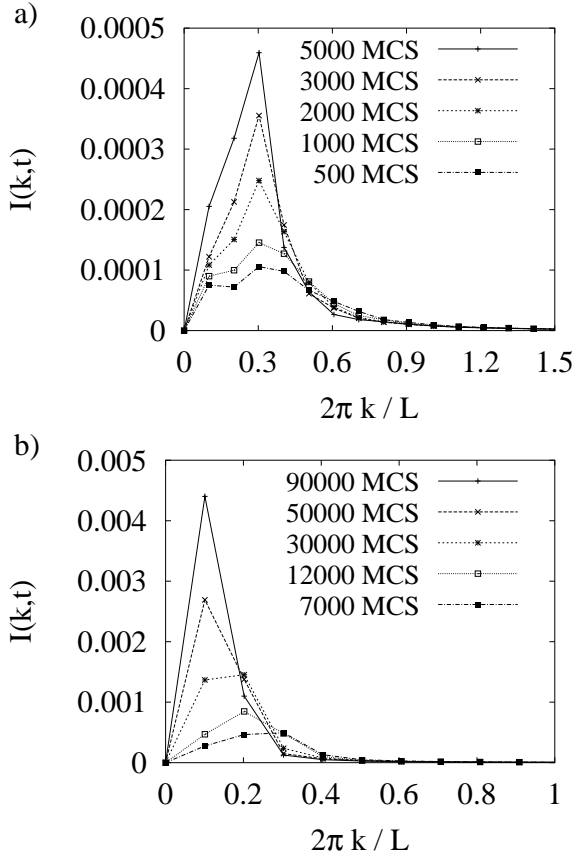,width=0.95\linewidth,angle=0}
  \end{center}
  \caption{Intermediate scattering function $I(k,t)$ for
    various times $t$. In the early time-regime a) the structure develops
    mainly at some fixed characteristic wavelength, while for later
    times b) the peak position of $I(k,t)$ shifts to smaller
    values of $k$ demonstrating coarsening of the domain patterns. The
    function is only evaluated at the physically meaningful integer
    values of $k$. The data are determined for a system of $M=4000$
    particles of size $N=50$.}
  \label{fig:Spektrum}
\end{figure}
To study the spontaneous phase separation in the GEM we consider an
equilibrated symmetric blend ($f_A\!=\!0.5$) for $\delta\!=\!0$ and
increase the mismatch parameter instantaneously to $\delta\!=\!0.2$ at
time $t\!=\!0$.  The power spectrum of the concentration fluctuations
is given by the intermediate scattering function 
\begin{equation}
I(k,t) =
\langle\rho_A(\textbf{k},t) \rho_A(-\textbf{k},t)\rangle-c_A^2\,
\delta(\textbf{k})\,,
\label{eq:intmed}
\end{equation}
where $\rho_A(\textbf{k},t)$ is the Fourier transform of the total monomer
density $\varrho_A'(\textbf{y},t)\!=\!\sum_{i=1}^{M_A}
\varrho_{A,i}'(\textbf{y},t)$ of $A$-ellipsoids. The power spectrum depends on
$k\!=\!|\textbf{k}|$ only and an explicit expression of $I(k,t)$ in terms of
particle positions, orientations and sizes is derived in
appendix~\ref{app:ft}.

\begin{figure}[!b]
  \begin{center}
    \epsfig{file=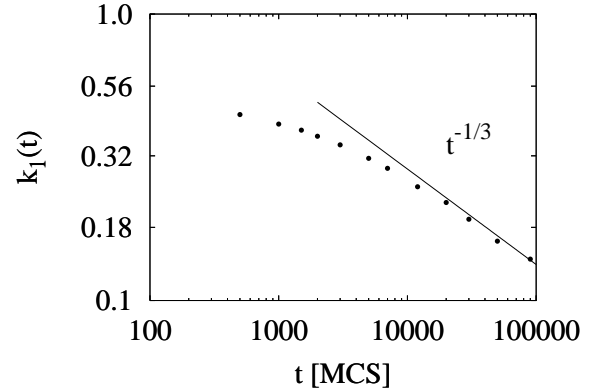,width=0.9\linewidth,angle=0}
  \end{center}
  \caption{Time-dependence of the first moment of the intermediate
    scattering function measuring the inverse domain size in
    double-logarithmic representation. At late stages the asymptotic
    Lifshitz-Slyozov behavior $k_{1}(t) \sim t^{-1/3}$ is reached.}
  \label{fig:first.moment}
\end{figure}

The results for $I(k,t)$ are presented in Fig.~\ref{fig:Spektrum}. At early
times a peak develops in $I(k,t)$, which is associated with a characteristic
demixing length. For $t\!\le\!5000$ MCS the peak position is almost constant
in time. At later times the domain pattern begins to coarsen and the peak of
$I(k,t)$ shifts to smaller values of $k$.  To quantify this coarsening process
we consider the normalized first moment
\begin{equation}
k_1(t)= \frac{\int_{0}^{\infty} dk\,k\,I(k,t)}
                  {\int_{0}^{\infty} dk\,I(k,t)}
\end{equation}
of the intermediate scattering function, which is a measure of the inverse
domain size. As shown in Fig.~\ref{fig:first.moment} the results point to an
asymptotic approach towards the Lifshitz-Slyozov growth law
$k_1(t)\!\sim\!t^{-1/3}$. We note that in more detailed microscopic models it
is hard to reach the asymptotic coarsening regime due to the slow diffusion of
the chain molecules (see the discussion in Sec.~\ref{sec:melt}).
Nevertheless, in some simulations of such models the Lifshitz-Slyozov growth
law could be seen \cite{Sariban/Binder:1991,Brown/Chakrabarti:1993}.

%% file: summary.cm.tex
\section{Concluding Remarks}
In summary we have shown that the GEM provides a useful tool to study polymer
systems on semi-macroscopic time and length scales.  The input quantities,
that are the probability $P(\textbf{S})$ for an ellipsoid to have a shape
$\textbf{S}$ and the mean monomer density $\varrho(\textbf{x};\textbf{S})$ in
the principal axis system for a given shape, were investigated in detail for
Gaussian chains and expressed in terms of approximate formulae. Both
homogeneous systems and binary mixtures have been simulated on the basis of
these input quantities by means of a discrete Monte Carlo procedure. The
results from these simulations were shown to reproduce the basic scaling
relations of polymer physics.

A valuable feature of the GEM is that the interactions between the
monomers of the same ellipsoid and different ellipsoids can be treated
on the same footing by including self-interactions as discussed in
Sec.~\ref{subsec:selfinter}. As a consequence, no tuning of the
interaction parameter $\hat\epsilon$ with $N$ is necessary to obtain
the desired Gauss type behavior in dense systems. On the other hand,
an important motivation of the work of Murat and Kremer
\cite{Murat/Kremer:1998} for setting up the ellipsoid model was to
keep a direct link to more detailed microscopic models. For such
underlying models it seems that a tuning of $\hat\epsilon$ with $N$
can not be avoided.  However, one may ask, if it is really necessary
to compare many particle simulations of the underlying microscopic
model with many particle simulations of the associated ellipsoid model
in order to establish the dependence of $\hat\epsilon$ on $N$, as it
was done in \cite{Murat/Kremer:1998}. An alternative route, dealing
only with the single chain behavior of the underlying microscopic
model, would be to use the Gaussian form for $P(\textbf{S},N)$ from
eq.~(\ref{eq:prod.PS}), where $N$ is the number of Kuhn segments of
the real microscopic chain with $N_{\rm mic}$ monomers.  Then one
should include self-interactions (see eq.~(\ref{eq:Fintersum})) with
$\hat\epsilon\!=\!\hat\epsilon(N)$ in the associated ellipsoid model
in such a manner that the true behavior of $P(\textbf{S},N_{\rm mic})$
for single chains of the microscopic model is best accounted for,
e.g.\ by requiring the first moment of $P_R(R_G^2,N)$ of the ellipsoid
model with self-interaction to equal the first moment of the true
$P_R(R_G^2,N_{\rm mic})$. In this way a later tuning of $\hat\epsilon$
with $N$ in the many particle simulation should no longer be
necessary. The procedure seems to be particularly promising, if
effective interactions of longer range are included in
eq.(\ref{eq:Finterv}), which mirror the interactions of the monomers
in the underlying microscopic model.

As outlined in the Introduction, our motivation for studying the GEM stems
largely from our wish to study the behavior of thin films of polymer mixtures.
In this respect we regard the work for the bulk behavior outlined here as a
suitable basis for further investigation. The important influence of substrate
interactions on the demixing kinetics of thin films can be easily taken into
account by incorporating the coupling of the monomer density to an external
potential in the free energy functional. Free surfaces can be modeled by
introducing attractive interactions between the centers of the ellipsoids.
These issues will be addressed in the near future.

\acknowledgements

We should like to thank W.~Dieterich and J.~Baschnagel for very
interesting discussions and B.~Rinn for supplying a parallelization
package. Financial support by the Deutsche Forschungsgemeinschaft (SFB
513, Ma~1636/2-1) is gratefully acknowledged.


%% file: appendix.cm.tex
\begin{appendix}

\section{Parameters of the distribution functions}
\label{app:constants} 
The determination of the parameters is based on the fact that
the moments of the heuristic functions are calculable analytically.
Therefore we define the constants $a_{R}$ and $d_{R}$ for
$P_{R}(R_{G}^{2})$ in such a way, that the first two moments $\langle
R_{G}^{2} \rangle$ and $\langle R_{G}^{4} \rangle$ are exact in the
limit $N \rightarrow \infty$. With this definition and the well known
\cite{Yamakawa:1971} results $\langle R_{G}^{2} \rangle \sim N/6$ and
$\langle R_{G}^{4} \rangle \sim 19\,N^{2}/540$ we are led to the two
equations
\begin{equation}
  a_{R} = \frac{1}{6} \frac{K_{0}(2 d_{R})}{d_{R} K_{1}(2 d_{R})}\;,
  \quad  \frac{K_{0}(2 d_{R}) K_{2}(2 d_{R})}{36\, K_{1}^{2}(2 d_{R})}
  = \frac{19}{540} \;.
\end{equation}
At this point we also give a more quantitative test of
$P_{R}(R_{G}^{2})$ by comparison of the higher moments of $R_{G}^{2}$.
The asymptotic exact result for the third moment is $\langle R_{G}^{6}
\rangle/N^{3} = 631/68040 \approx 0.009274$, while we find $\langle
R_{G}^{6} \rangle/N^{3} = 0.009280$. For the fourth moment the exact
result $\langle R_{G}^{8} \rangle/N^{4} = 1219/408240 \approx
0.002986$ has to be compared with $\langle R_{G}^{8} \rangle/N^{4} =
0.003000$.

For $P_{\alpha}(S_{\alpha})$ the same procedure leads to 
\begin{eqnarray}
\frac{\langle S_{\alpha} \rangle}{d_{\alpha}}
\frac{K_{n_{\alpha}-1}(2 d_{\alpha})}{K_{n_{\alpha}-2}(2 d_{\alpha})} 
&=& a_{\alpha} \;,\\
\frac{K_{n_{\alpha}-3}(2 d_{\alpha}) K_{n_{\alpha}-1}(2 d_{\alpha})}
{K_{n_{\alpha}-2}^{2}(2 d_{\alpha})} 
&=&\frac{\langle
  S_{\alpha}^{2}\rangle}{\langle S_{\alpha} \rangle^{2}}\;.
\end{eqnarray}
Since no exact analytical result for the moments $\langle
S_{\alpha}^{m} \rangle$ exists to our knowledge, these quantities are
determined by Monte-Carlo simulations of Gaussian chains of length
$N=1000$. The results are given in Tab.~\ref{tab:Sa.moments}.

\section{Improved formula for the monomer density}
\label{app:better.rhoy}
For the sake of completeness, we present an improved heuristic formula
for the density of the $x_{2}$-axis:
\begin{eqnarray}
\label{eq:better.rhoy}
  \tilde \varrho_2(u_2)=\frac{1}{\sqrt{2 \pi} 2\,\sigma'_2}&&
  \left[ \exp \left( -\frac{(u_2-\bar u_2)^2}{2\,{\sigma'_2}^{2} }\right)
    + \right.\nonumber \\ 
&& \hspace*{0.2cm}\left. \exp\left( -\frac{(u_2+\bar
  u_2)^2}{2\,{\sigma'_2}^{2}}\right)\right] 
\end{eqnarray}
Least squares fit of the constants yield $\bar u_{2} = 0.674$ and
$\sigma'_2 = 0.748$.

\section{Overlap integrals}
\label{app:analytic.integration}
When inserting eqs.(\ref{eq:rhoa.heuristic},\ref{eq:varrho})
into eqs.(\ref{eq:trafo},\ref{eq:Finter}) the calculation of
$F_{\text{inter}}$ reduces to a sum of nine terms of the form
\begin{equation}
  \label{eq:overlap.integral}
  \int_{-\infty}^{+\infty}\hspace*{-0.2cm} d^3y\,\,
  \chi^{(1)}(\textbf{y})\,\chi^{(2)}(\textbf{y})\,,
\end{equation}
where the functions $\chi^{(j)}(\textbf{y})\;,\;j=1,2$ are given by 
\begin{eqnarray}
\label{eq:chidef}
  \chi^{(j)}(\textbf{y})&=&\frac{1}{(2\pi)^{3/2} \prod_{\alpha=1}^3
   \sigma_{\alpha}^{(j)}} \exp\left(-\frac{1}{2}\sum_{\alpha}
   \left(\frac{x_{\alpha}^{(j)}(\textbf{y})}{\sigma_{\alpha}^{(j)}} \right)^{2}
   \right)\,,\nonumber\\[0.2cm]
&&  \hspace*{-1cm}
x_{\alpha}^{(j)}(\textbf{y})=\sum_{\beta} {\cal R}_{\alpha \beta}^{(j)}
  (y_{\beta}-w_{\beta}^{(j)}) \,, \hspace*{0.3cm}
\sigma_{\alpha}^{(j)} = R_{\alpha}^{(j)}\sigma_{\alpha}\,.
\end{eqnarray}
The $\sigma_{\alpha}$ were defined in eq.~(\ref{eq:rhoa.heuristic}) and
specified in Tab.~\ref{tab:rho.constants}. Moreover,
$w_\alpha^{(j)}\!=\!\pm{\cal R}_{1,\alpha} R_1^{(j)}
u_0+r_\alpha^{(j)}$ or $w_\alpha^{(j)}\!=\!r_\alpha^{(j)}$ depending
on which of the three terms in eq.~(\ref{eq:rhoa.heuristic-a})
contributes to $\chi^{(j)}(\textbf{y})$. We define
\begin{eqnarray}
\sigma&\equiv&\prod_{\alpha=1}^3 \sigma_{\alpha}^{(1)} \sigma_{\alpha}^{(2)}\\
T(\textbf{y})&\equiv&\sum_{\alpha=1}^3\left\{
  \left(\frac{x_{\alpha}^{(1)}(\textbf{y})}{\sigma_{\alpha}^{(1)}}
  \right)^{2}
 +\left(\frac{x_{\alpha}^{(2)}(\textbf{y})}{\sigma_{\alpha}^{(2)}}
  \right)^{2}\right\}\;. 
\end{eqnarray}
With these definitions the expression (\ref{eq:overlap.integral}) reads
\begin{equation}
  \label{eq:overlap.integral2}
  \frac{1}{(2\pi)^{3} \sigma} \int_{-\infty}^{+\infty}\hspace*{-0.2cm} d^3y\,
  \exp\left[-\frac{1}{2}\,T(\textbf{y})\right]\,.
\end{equation}
Since
\begin{eqnarray}
\sum_{\gamma}\left(\frac{x_{\gamma}^{(j)}(\textbf{y})}
  {\sigma_{\gamma}^{(j)}}\right)^{2}&=& 
  \sum_{\alpha,\beta}{\cal L}_{\alpha \beta}^{(j)} 
\left(y_{\alpha}y_{\beta}\!-\!2w_{\alpha}^{(j)}y_{\beta}\!+\!w_{\alpha}^{(j)}
  w_{\beta}^{(j)}\right)\,,\nonumber\\
{\cal L}_{\alpha \beta}^{(j)}&=& \sum_{\gamma} \frac{{\cal R}_{\gamma
  \alpha}^{(j)}{\cal R}_{\gamma \beta}^{(j)}}
  {\left(\sigma_{\gamma}^{(j)}\right)^{2}}\,,  
\end{eqnarray}
it follows that
\begin{equation}
  \label{eq:T2}
  T(\textbf{y}) = \sum_{\alpha,\beta} {\cal A}_{\alpha \beta}
  y_{\alpha} y_{\beta} + \sum_{\alpha} B_{\alpha} y_{\alpha} + C \;, 
\end{equation}
where
\begin{eqnarray}
  {\cal A}_{\alpha \beta}&=& 
  {\cal L}_{\alpha \beta}^{(1)} + {\cal L}_{\alpha \beta}^{(2)}\,,\\ 
  B_{\alpha}  &=& -2 \sum_{\beta} \left( {\cal L}_{\beta \alpha}^{(1)}
  w_{\beta}^{(1)} + {\cal L}_{\beta \alpha}^{(2)} w_{\beta}^{(2)}
  \right)\,,\\ 
  C &=& \sum_{\alpha,\beta}\left( {\cal L}_{\alpha \beta}^{(1)}
  w_{\alpha}^{(1)} w_{\beta}^{(1)} + {\cal L}_{\alpha
  \beta}^{(2)} w_{\alpha}^{(2)} w_{\beta}^{(2)} \right)\,.
\end{eqnarray}
The quadratic form (\ref{eq:T2}) can be rewritten as
\begin{equation}
  T(\textbf{y})=\sum_{\alpha,\beta}
  (y_{\alpha}-w_{\alpha})\, {\cal D}_{\alpha \beta} \,
  (y_{\beta}-w_{\beta}) + E \,,
\end{equation}
which yields
\begin{equation}
  \int_{-\infty}^{+\infty}\hspace*{-0.2cm}d^3y\,
  \chi^{(1)}(\textbf{y})\,\chi^{(2)}(\textbf{y}) =
  \frac{\exp(-E/2)}{(2\pi)^{3/2}\sigma\sqrt{{\rm det}{\,\cal D}}}\,,
\end{equation}
\begin{eqnarray}
  4\,{\rm det}{\,\cal D}&=&4\,{\cal A}_{11}{\cal A}_{22}{\cal A}_{33} + 
{\cal A}_{12}{\cal A}_{13}{\cal A}_{23}\nonumber\\
&&\hspace*{0cm} - {\cal A}_{13}^{2}{\cal A}_{22} - 
{\cal A}_{23}^{2}{\cal A}_{11} - {\cal A}_{12}^{2}{\cal A}_{33}\,,
\end{eqnarray}
\begin{eqnarray}
E=C+\frac{1}{16\,{\rm det}{\,\cal D}} \Bigl[&&
        B_1^{2} ({\cal A}_{23}^{2}-4\,{\cal A}_{22}{\cal A}_{33})\\ 
{}+&&B_{2}^{2} ({\cal A}_{13}^{2}-4\,{\cal A}_{11}{\cal A}_{33})\nonumber\\ 
{}+&&B_{3}^{2} ({\cal A}_{12}^{2}-4\,{\cal A}_{11}{\cal A}_{33}) \nonumber\\ 
{}+&&2\,B_{1}B_{2} (2\,{\cal A}_{12}{\cal A}_{33}-
{\cal A}_{13}{\cal A}_{23}) \nonumber\\
{}+&&2\,B_{1}B_{3} (2\,{\cal A}_{13}{\cal A}_{22}-
{\cal A}_{12}{\cal A}_{23}) \nonumber\\ 
{}+&&2\,B_{2}B_{3} (2\,{\cal A}_{23}{\cal A}_{11}
-{\cal A}_{12}{\cal A}_{13}) \Bigr]\,.\nonumber
\end{eqnarray}

\section{Fourier transform of the monomer density}
\label{app:ft}
For a total monomer density defined by 
\begin{equation}
  \varrho'(\textbf{y}) \equiv \frac{1}{M(N\!+\!1)} \sum_{i=1}^{M}
  \varrho_{i}'(\textbf{y})\,,
\end{equation}
the Fourier transform is given by
\begin{equation}
  \rho(\textbf{k}) = \int_{-\infty}^{\infty}\hspace*{-0.2cm}
  d^{3}y\,\varrho'(\textbf{y})\,\exp(-i\,\textbf{k}\!\cdot\!\textbf{y})\,.
\end{equation}
Inserting the $\varrho_{i}'(\textbf{y})$ from eq.~(\ref{eq:trafo})
with (\ref{eq:rhoa.heuristic},\ref{eq:varrho}) yields
\begin{eqnarray}
  \rho(\textbf{k})&=&\frac{1}{2+c_{1}} \frac{1}{M} \sum_{i=1}^{M} 
  \left\{ \int_{-\infty}^{\infty} \hspace*{-0.2cm} d^{3}y\, 
      e^{-i\,\textbf{k}\cdot\textbf{y}}  \nonumber \right.\\ 
&& \left.  \phantom{\int_{-\infty}^{\infty} \hspace*{-0.2cm} d^{3}y\, 
      e^{-i\,\textbf{k}\cdot\textbf{y}}} \hspace{-1.5cm}
    \times \Bigl[ \chi_{\text{a}}^{(i)}(\textbf{y}) +
      \chi_{\text{b}}^{(i)}(\textbf{y}) +
  c_{1}\,\chi_{\text{c}}^{(i)}(\textbf{y})   \Bigr] \right\} \,.
\end{eqnarray}
The three different functions
$\chi_{\text{a}}^{(i)}(\textbf{y})$, $\chi_{\text{b}}^{(i)}(\textbf{y})$ and
$\chi_{\text{c}}^{(i)}(\textbf{y})$ are defined in 
eq.~(\ref{eq:chidef}), where all different kinds of $w_{\alpha}^{(j)}$
occur, i.e. $w_{\alpha,\text{a}}^{(j)}\!=\!+\!{\cal R}_{1,\alpha} R_1^{(j)}
u_0+r_\alpha^{(j)}$, $w_{\alpha,\text{b}}^{(j)}\!=\!-\!{\cal R}_{1,\alpha}
R_1^{(j)} u_0+r_\alpha^{(j)}$ and
$w_{\alpha,\text{c}}^{(j)}\!=\!r_\alpha^{(j)}$, respectively. 
The integration is straightforward yielding the final result
\begin{eqnarray}
  \rho(\textbf{k})&=&\frac{1}{(2+c_{1})
    \sigma_{1}\sigma_{2}\sigma_{3}} \frac{1}{M} \sum_{i=1}^{M}
  \frac{1}{\sqrt{S_{1}^{(i)}S_{2}^{(i)}S_{3}^{(i)}}}
  \frac{1}{\sqrt{{\rm det}{\,\cal G}^{(i)}}} \nonumber\\
&&\hspace*{-0.5cm}{}\times\exp
\left(-\frac{1}{2} \sum_{\alpha, \beta} k_{\alpha}
    {\cal G}_{\alpha\beta}^{(i)^{-1}} k_{\beta} \right) \exp
  \left(-i\,\sum_{\beta} k_{\beta}r_{\beta}^{(i)} \right)\nonumber\\
&&\hspace*{-0.5cm}{}\times\left[ 2\, \cos \left( \sum_{\beta} k_{\beta}
      {\cal R}_{1\beta}^{(i)} R_{1}^{(i)} u_{0} \right) +c_{1} \right]
\label{eq:rhok}
\end{eqnarray}
Here we defined ${\cal G}_{\alpha\beta}^{(i)}=\sum_{\gamma} g_{\gamma}^{(i)}
{\cal R}_{\gamma\alpha}^{(i)}{\cal R}_{\gamma\beta}^{(i)}$ and
$g_{\gamma}^{(i)}=\left( \left(\sigma_{1}^{2}S_{1}^{(i)}\right)^{-1},
  \left(\sigma_{2}^{2}S_{2}^{(i)}\right)^{-1},
  \left(\sigma_{3}^{2}S_{3}^{(i)}\right)^{-1} \right)$. The center of
mass of the $i$'th particle is denoted by $r_{\beta}^{(i)}$.
The intermediate scattering function is then obtained by using
eq.~(\ref{eq:rhok}) in eq.~(\ref{eq:intmed}) for all times $t$ (for
the A ellipsoids).

\end{appendix}


%% file: references.cm.tex

%% file: sem.methoden.cm.bbl
\begin{thebibliography}{99}
\bibitem{small.N} We examined chains of length $N = 2-7$ and $10$,
  too, and found larger deviations from scaling.
\end{thebibliography}
